\documentclass[a4paper,11pt]{article}

\pdfoutput=1 

\usepackage{jcappub} 

\usepackage[T1]{fontenc} 

\usepackage{subcaption}

\title{\boldmath Does Planck Actually ``See'' the Bunch-Davies State?}

\author[a]{Rose Baunach,}
\author[a]{Nadia Bolis,}
\author[b,1]{R. Holman,\note{Corresponding author.}}
\author[c]{Stacie Moltner,}
\author[b]{Benoit J. Richard}

\affiliation[a]{Center for Quantum Mathematics and Physics and Department of Physics and Astronomy, University of California at Davis, One Shields Ave, Davis, CA 95616, USA}
\affiliation[b]{Minerva Schools at KGI, 1145 Market Street, San Francisco, CA 94103, USA}
\affiliation[c]{Theory Group, Department of Physics, The University of Texas at Austin, Austin, TX 78712, USA}

\emailAdd{baunach@ucdavis.edu}
\emailAdd{nbolis@ucdavis.edu}
\emailAdd{rholman@minerva.kgi.edu}
\emailAdd{staciemoltner@utexas.edu}
\emailAdd{brichard@minerva.kgi.edu}

\abstract{To what extent can the Planck satellite observations be interpreted as confirmation of the quantum part of the inflationary paradigm? Has it ``seen'' the Bunch-Davies state? We compare and contrast the Bunch-Davies interpretation with one using a so-called \emph{entangled} state in which the fluctuations of a spectator scalar field are entangled with those of the metric perturbations $\zeta$. We first show how a spectator scalar field $\Sigma$, with an expectation value $\sigma(t)$ that evolves in time, will generically generate such a state. We then use this state to compute the power spectrum $P_{\zeta}(k)$ and thence the temperature anisotropies $C_l$ in the Cosmic Microwave Background (CMB). We find interesting differences from the standard calculations using the Bunch-Davies (BD) state. We argue that existing data may already be used to place interesting bounds on this class of deviations from the BD state and that, for some values of the parameters of the state, the power spectra may be consistent with the Planck satellite data.}

\begin{document}
\maketitle
\flushbottom

\section{Is Bunch-Davies All There Is?}
\label{sec:intro}
The inflationary paradigm~\cite{Guth:1980zm, Starobinsky:1980te, Kazanas:1980tx, Albrecht:1982wi, Linde:1981mu} can be thought of as comprising two parts. The first is concerned with {\em models} of inflation, that is in finding viable field theoretic realizations of inflation. The second, which is the focus of this work, deals with the quantum mechanics of inflationary perturbations.

One of the signal successes of inflation was the realization that quantum fluctuations during the inflationary phase could be stretched to cosmological length scales and that they would decohere so as to be able to be treated classically and serve as a causally generated source of density fluctuations. These in turn would drive the formation of cosmic structure in the early universe~\cite{Guth:1982ec, Bardeen:1983qw, Hawking:1982cz}. In order to calculate the power spectrum of these fluctuations and how these might show up in physical observables such as temperature anisotropies in the Cosmic Microwave Background (CMB), we need to know the quantum state of the field representing the metric perturbations. 

For a scalar field in a near de Sitter background spacetime, there is a preferred quantum state, the so-called Bunch-Davies (BD) state~\cite{Bunch:1978yq}. Though the notion of a lowest energy state is of dubious value in a dynamical spacetime, the BD state has a number of ``ground state'' traits. It is a state of maximal symmetry in that it is invariant under the symmetries of de Sitter space and it is an adiabatic state~\cite{BirrellDavies1982}, which is the nearest approximation to a state devoid of particles that can be obtained in this context. In fact, it is the state that in the short-distance, short-time limit approaches the Minkowski space vacuum state for a scalar field theory. 

From this perspective, the BD state becomes a natural one to use for the computation of inflationary cosmological observables. Given a model of inflation, using this state allows us to make predictions about various aspects of the CMB power spectrum, bi-spectrum, as well as other cosmological observables. It is thus a linchpin of the inflationary paradigm. But is the BD state the true state of the inflaton? How can we tell? 

If we {\em were} to find that the BD state is necessarily the quantum state of the inflaton, this would bring up a number of questions, not least of which is: What makes the BD state so stable with respect to all the potentially non-adiabatic physical effects that would most certainly be taking place before or at the onset of inflation? Alternatively, if we find other states that could conceivably fit the bill as consistent inflationary quantum states, then the structure of these states might provide hints to pre-inflationary physics.

These questions drive us to explore the issue of how to delineate the space of allowable inflationary states. In general, this task is a difficult one, made more so by the paucity of cosmological probes that can be directly brought to bear on it. However, there are some requirements that an inflationary quantum state must satisfy. First and foremost, {\em it must allow inflation to occur!} This is to be interpreted as the requirement that the expectation value of the stress tensor of the inflaton should not give rise to an energy density that exceeds that coming from the inflaton potential. Second, it should give rise to values of cosmological observables consistent with those measured by probes of the CMB as well as those of large scale structure (LSS). While these two conditions are the {\it sine qua non} of any potential inflaton quantum state, we will add another restriction so as to make the problem tractable; we will assume that the state is Gaussian in the field fluctuations. While this appears to just lead to the standard free field theory quantum state, we will also allow for the existence of {\em spectator} scalar fields and for the possibility that the state describing spectator field fluctuations is entangled (in the sense of refs.~\cite{Albrecht:2014aga,Bolis:2016vas,Collins:2016ahj,Bolis:2019fmq}) with the state of the metric perturbations. This opens up the space of states to a larger set than just the free field Gaussian state of metric fluctuations, albeit, in a way that is still amenable to analysis. 

Thus, we will assume that an inflationary period is induced by the slow-roll of an inflaton field $\Phi$. We will also suppose that the spectator field does not couple to the inflaton directly so that the potential $V(\Phi, \Sigma)=V_{\rm inf}(\Phi)+V_{\rm spec}(\Sigma)$. We will work in the comoving gauge in which the fluctuations $\delta \phi$ of the inflaton around its rolling expectation value $\phi(t)=\langle \Phi(\vec{x},t)\rangle$ are gauged away and all their information is encoded in the scalar metric perturbation $\zeta$ (which, on superhorizon scales, is proportional to the curvature perturbation ${\mathcal R}$).

In the previous work\footnote{Note that there is also a large body of work that investigates deviations from the Bunch-Davies state in a non-entangled context, e.g.~\cite{Martin_2001,Danielsson_2002, Kaloper_2002, Collins_2005, Sriramkumar_2005, Collins_2006, Albrecht_2011, Kundu_2012, Dey_2012, Carney_2012, Agarwal_2013, Kundu_2014, Ashoorioon_20141, Ashoorioon_20142, DiTucci:2019xcr}.} done with these states, either the initial entanglement was assumed to be non-zero, but otherwise arbitrary, or it was generated~\cite{Holman:2019spa} by looking at cubic order terms in the $\zeta$-scalar action~\cite{Weinberg:2005vy, delRio:2018vrj}. In this work, we will show that, in fact, non-trivial entanglement can be generated even at the quadratic level of the coupled $\zeta$-scalar system, as long as the expectation value of the scalar is initially displaced from the minimum of its potential. Given the appearance of (light) scalar fields in a number of extensions of the standard model as well as the existence of the Higgs, it could be argued that our entangled states might in fact be a generic outcome of early universe evolution.

This then is the aim of our paper: to show that the addition of a spectator field can generate a non-trivial entanglement between the metric perturbations and scalar field fluctuations and then to compute the power spectrum and use it to construct the CMB temperature anisotropies. We will consider both a free massive field as well as a field with an axion-type potential as a spectator field.

What we find is that in both cases, as long as the scalar has an expectation value that is either displaced from its minimum and/or has a non-zero time derivative, a nontrivial entangled state will be generated. The resulting power spectra depend on the initial values of the scalar field and its time derivative, as well as the ratio of the scalar's mass relative to the Hubble parameter of the de Sitter expansion. They exhibit a variety of behaviors, depending on the parameter values. Interestingly, even in the cases where the primordial power spectrum deviates significantly from that of the Bunch-Davies case, when the entanglement happens near when the largest length scales appearing in the CMB leave the inflationary horizon, the $C_l$s can essentially remain unchanged from the non-entangled case. However, for other parameter values, we will see that large deviations can occur.

Our interest in the quantum state of the system---as opposed to the observables of interest in collider physics such as S-matrix elements, scattering amplitudes and the like---dictates the technique we will use in this work. Schrödinger picture field theory~\cite{Boyanovsky:1993xf,Freese:1984dv} is the natural tool in this setting, and we review its use in determining the inflaton quantum state in the next section. In section~\ref{sec:powerspectrum}, we compute the $\zeta$ power spectrum and the $C_l$s for a variety of initial conditions for the rolling expectation value of the spectator scalar. Finally, we discuss our results and further research directions in section~\ref{sec:conclusions}.

\section{The Schrödinger Wave Functional for the $\zeta$-Spectator Scalar System}
\label{sec:schroFT}

We have discussed the use of Schrödinger field theory at length in previous work on entangled states~\cite{Albrecht:2014aga,Bolis:2016vas,Bolis:2019fmq}, so we will settle for a quick summary here. As mentioned in the introduction, we are {\em not} concerned with observables such as scattering amplitudes, but rather with the form and evolution of the quantum state itself. In the Schrödinger picture this corresponds to the construction of a wavefunctional depending on the relevant field configurations defined on the spatial hypersurface at conformal time $\eta$, $\Psi\left[\zeta(\cdot), \Sigma(\cdot); \eta\right]$. This wavefunctional then satisfies the Schrödinger equation
\begin{equation}
\label{eq:Seqn}
i\partial_{\eta} \Psi\left[\zeta(\cdot), \Sigma(\cdot); \eta\right]=H\left[\Pi_{\zeta}, \Pi_{\Sigma}, \zeta, \Sigma;\eta\right] \Psi\left[\zeta(\cdot), \Sigma(\cdot); \eta\right],
\end{equation}
where $\Pi_{\zeta}, \Pi_{\Sigma}$ are the canonically conjugate momenta to $\zeta$ and $\Sigma$, respectively. In equation \eqref{eq:Seqn}, we have included the explicit time dependence of the Hamiltonian coming from the expansion of the Universe as encoded in the scale factor $a(\eta)$. Given the wavefunctional, cosmological observables are simply expectation values of the relevant operators, taken in the Schrödinger picture. 

\subsection{Constructing the $\zeta$-$\Sigma$ Hamiltonian}
\label{subsec:zetasigma}
As mentioned above, the system we will consider is one where the scalar metric perturbations $\zeta$ are entangled with a scalar field $\Sigma$. Before we proceed with the calculation, though, it is worth taking the time to understand some physics details. We need the spectator field to truly be a spectator. What that means is that we have to ensure that the energy density in $\Sigma$ must be far smaller than that of the inflaton. This allows us to neglect the effects of the isocurvature perturbations induced by $\Sigma$, at least to lowest order. Including such effects will be left to later work. 

In order to begin, we need the action for the scalar metric fluctuations $\zeta$ coupled to a scalar field $\Sigma$ whose potential $V(\Sigma)$ drives its dynamics. We take $\langle \Sigma\rangle\equiv \sigma(\eta)$, where the expectation value is taken with the wavefunctional $\Psi$. Arriving at this action entails using the ADM~\cite{Arnowitt:1962hi} form of the Einstein action, writing the metric in terms of the lapse and shift functions, $N,\ N_i$ respectively, as well as $\zeta$, and then solving the constraint equations for the Lagrange multipliers $N,\ N_i$. The fact that $\Sigma$ has a non-trivial expectation value complicates matters somewhat, and we use the Mathematica package $MathGR$~\cite{Wang:2013mea} to aid us in our calculations. Also, we take advantage of the spatial flatness of the constant-$\eta$ hypersurfaces and write the action directly in terms of the momentum space modes $\zeta_{\vec{k}}$ and $\chi_{\vec{k}}$, where we expand the field $\Sigma$ about its expectation value: $\Sigma(\vec{x},\eta) = \sigma(\eta)+\chi(\vec{x},\eta)$. 

We will only keep terms in the action out to quadratic order in the fluctuations $\zeta_{\vec{k}}$ and $\chi_{\vec{k}}$, reasoning that this will suffice in order to set up the Schrödinger equation for a Gaussian state. 

There is a term that is independent of the fluctuations but only depends on the background cosmology; we will neglect this term since we can rephase the wavefunctional to absorb its effects. In addition, there exists a term linear in $\chi$ that contributes to the tadpole; it's proportional to the zero mode equation of motion, which, at the classical level is given by:
\begin{equation}
\label{eq:zeromodeclass}
\sigma^{\prime \prime}(\eta)+ 2 \frac{a^{\prime}(\eta)}{a(\eta)} \sigma^{\prime}(\eta) +a^2(\eta)\partial_{\sigma}V(\sigma)=0.
\end{equation}
We will assume this equation is satisfied, so we can neglect this term in the action and hence in the Hamiltonian (see~\cite{Holman:2019spa} for how higher order corrections to the zero mode equation can be implemented in the presence of entanglement). 

Doing all of this, we find the action to quadratic order is given by:
\begin{eqnarray}
\label{eq:kspaceaction}
&& S=\int d\eta\ \int \frac{d^3 k}{(2\pi)^3}\ {\mathcal L}_k\nonumber\\
&& {\mathcal L}_k = a^2(\eta)\left\{\frac{1}{2} \vec{X}^{T \prime}_{\vec{k}}\ {\mathcal O}\ \vec{X}^{\prime}_{-\vec{k}}+\vec{X}^{T \prime}_{\vec{k}} \ {\mathcal M}\ \vec{X}_{-\vec{k}}-\frac{1}{2} \vec{X}^T_{\vec{k}}\ \Omega_k^2\ \vec{X}_{-\vec{k}}\right\}, 
\end{eqnarray}
where primes denote conformal time derivatives,
$$
 \vec{X}_{\vec{k}}= \begin{pmatrix}
    \zeta_{\vec{k}}\\ 
    \chi_{\vec{k}}
  \end{pmatrix},
$$
and the matrices ${\mathcal O}$, ${\mathcal M}$, and $\Omega_k^2$ are given by:
\begin{subequations}\label{eq:actionmatrices}
\begin{align}
\label{eq:actionmatrices:1}
{\mathcal O} & = \begin{pmatrix}
	2 M_P^2\ \epsilon & \ -\frac{\sigma^{\prime}}{{\mathcal H}}\\
	-\frac{\sigma^{\prime}}{{\mathcal H}} &\ 1
	\end{pmatrix}
\\
\label{eq:actionmatrices:2}
{\mathcal M} & = \begin{pmatrix}
	0 &\ \ \epsilon \sigma^{\prime}-\frac{a^2(\eta) \partial_{\sigma}V(\sigma)}{{\mathcal H}}\\
	3\sigma^{\prime} &\ 0
	\end{pmatrix}
\\
\label{eq:actionmatrices:3}
 \Omega^2 & =\begin{pmatrix}
	2 M_P^2\ \epsilon\ k^2 & 3 a^2(\eta) \partial_{\sigma}V(\sigma)-k^2 \frac{\sigma^{\prime}}{{\mathcal H}}\\
	3 a^2(\eta) \partial_{\sigma}V(\sigma)-k^2 \frac{\sigma^{\prime}}{{\mathcal H}} &\ \ \ k^2 +a^2(\eta) \partial^2_{\sigma}V(\sigma)+(3-\epsilon) \frac{\sigma^{\prime 2}}{M_P^2}+\frac{2 a^2(\eta) \sigma^{\prime} \partial_{\sigma}V(\sigma) }{M_P^2 {\mathcal H}} \\ 
	\end{pmatrix}.
\end{align}
\end{subequations}
We have also defined the conformal time Hubble parameter $ {\mathcal H}$ via
$$
 {\mathcal H}\equiv \frac{a^{\prime}(\eta)}{a(\eta)}, 
 $$
as well as the slow-roll parameter $\epsilon$, defined in conformal time by ${\mathcal H}^{\prime}= (1-\epsilon){\mathcal H}^2$.

We note the following facts. First, the dimensions of the entries in ${\mathcal O}$ as well as the other matrices are different; this arises due to the fact that in position space, $\zeta$ has mass dimension $0$ while $\chi$ has mass dimension $1$, which in turn means that $\zeta_{\vec{k}}$ and $\chi_{\vec{k}}$ have dimensions $-3$ and $-2$, respectively. Furthermore, the mixing matrix ${\mathcal M}$, which mixes the positions and velocities, vanishes when $\sigma$ rests at the minimum (or maximum) of the potential $V(\sigma)$.

More importantly though, we notice that the mixing term involving ${\mathcal M}$ is {\em not} hermitian. We can see this by noting that for a real field $\phi$, $\phi^*_{\vec{k}}=\phi_{-\vec{k}}$, so that, after a $\vec{k}\leftrightarrow -\vec{k}$ change of variable in the $k$ integral, we can write the relevant term in the action as $\vec{X}^{\dagger \prime}_{\vec{k}} \ {\mathcal M}\ \vec{X}_{\vec{k}}$. Taking the hermitian conjugate and noting that ${\mathcal M}$ is a real matrix, we have
\begin{subequations}\label{eq:nonherm}
\begin{equation}
\label{eq:nonherm:1}
a^2(\eta)\left(\vec{X}^{\dagger \prime}_{\vec{k}} \ {\mathcal M}\ \vec{X}_{\vec{k}}\right)^{\dagger}=a^2(\eta)\left(\vec{X}^{\dagger}_{\vec{k}} \ {\mathcal M}^T\ \vec{X}^{ \prime}_{\vec{k}}\right).
\end{equation}
Integrating by parts and discarding the surface term allows us to rewrite this as 
\begin{equation}
\label{eq:nonherm:2}
a^2(\eta)\left(\vec{X}^{\dagger}_{\vec{k}} \ {\mathcal M}^T\ \vec{X}^{ \prime}_{\vec{k}}\right)=-a^2(\eta)\left(\vec{X}^{\dagger \prime}_{\vec{k}} \ {\mathcal M}^T\ \vec{X}_{\vec{k}}\right)-\vec{X}^{\dagger}_{\vec{k}} \ \partial_{\eta}\left(a^2(\eta) {\mathcal M}^{T }\right)\ \vec{X}_{\vec{k}}.
\end{equation}
\end{subequations}

To make the action hermitian, we replace 
\begin{equation}
\label{eq:hermreplace}
\vec{X}^{T \prime}_{\vec{k}} \ {\mathcal M}\ \vec{X}_{-\vec{k}}\rightarrow \frac{1}{2}\left(\vec{X}^{T \prime}_{\vec{k}} \ {\mathcal M}\ \vec{X}_{-\vec{k}}+\left(\vec{X}^{\dagger \prime}_{\vec{k}} \ {\mathcal M}\ \vec{X}_{\vec{k}}\right)^{\dagger}\right). 
\end{equation}
From \eqref{eq:nonherm:2} we see that this is equivalent to the combined operations of (i) replacing ${\mathcal M}\rightarrow {\mathcal M}_A$ and (ii) shift $\Omega_k^2 \rightarrow \Omega_k^2+\partial_{\eta}\left(a^2(\eta) {\mathcal M}_S\right)\slash a^2(\eta)$, where $S,A$ denote the symmetric and antisymmetric parts of ${\mathcal M}$.

To construct the Hamiltonian, we start with the momenta conjugate to $\zeta_{\vec{k}},\ \chi_{\vec{k}}$:
\begin{equation}
\label{eq:canonicalmom}
\vec{\Pi}_{\vec{k}} = \frac{\delta {\mathcal L}_k}{\delta \vec{X}^{\prime}_{-\vec{k}}}=a^2(\eta)\left[{\mathcal O} \vec{X}^{\prime}_{\vec{k}}+{\mathcal M}_A \vec{X}_{\vec{k}}\right]\Rightarrow \vec{X}^{\prime}_{\vec{k}}=\frac{1}{a^2(\eta)}{\mathcal O}^{-1} \vec{\Pi}_{\vec{k}}-{\mathcal O}^{-1}{\mathcal M}_A\vec{X}_{\vec{k}}.
\end{equation}

From the definition of the momentum space Hamiltonian density $H_k$ as $H_k=\vec{\Pi}_{\vec{k}}^T \vec{X}^{\prime}_{-\vec{k}}-{\mathcal L}_k$ we have:
\begin{eqnarray}
\label{eq:Hamiltonian}
H_k &=&\frac{1}{2 a^2(\eta)} \vec{\Pi}^T_{\vec{k}}\ {\mathcal O}^{-1}\  \vec{\Pi}_{-\vec{k}}-\frac{3}{2} \vec{\Pi}^T_{\vec{k}}\ {\mathcal O}^{-1}{\mathcal M}_A\ \vec{X}_{-\vec{k}} + \frac{1}{2}\vec{X}^T_{\vec{k}}\ {\mathcal M}_A^T {\mathcal O}^{-1}\ \vec{\Pi}_{-\vec{k}} \nonumber\\
&+&\frac{a^2(\eta)}{2} \vec{X}^T_{\vec{k}} \left(\Omega_k^2+\frac{\partial_{\eta}\left(a^2(\eta) {\mathcal M}_S\right)}{a^2(\eta)}+{\mathcal M}_A^T  {\mathcal O}^{-1} {\mathcal M}_A\right)\vec{X}_{-\vec{k}}.
\end{eqnarray}
The full Hamiltonian is then a momentum space integral of $H_k$. We should note that the middle two terms in $H_k$ are quantum mechanically ambiguous due to operator ordering issues. We deal with this by means of the Weyl prescription: $\vec{\Pi}_a \vec{x}_b\rightarrow (\vec{\Pi}_a \vec{x}_b+ \vec{x}_b\vec{\Pi}_a)\slash 2 $, where $a,b=1,2$. Doing this and using the fact that ${\mathcal O}$ and hence ${\mathcal O}^{-1}$ is symmetric, we can write
\begin{eqnarray}
\label{eq:Hamiltonian2}
H_k&=&\frac{1}{2 a^2(\eta)} \vec{\Pi}^T_{\vec{k}}\ {\mathcal O}^{-1}\  \vec{\Pi}_{-\vec{k}}- \frac{1}{2} \left[{\mathcal O}^{-1}{\mathcal M}_A\right]_{a b}\left(\vec{\Pi}_{ \vec{k} a} \vec{X}_{ -\vec{k} b} +\vec{X}_{ -\vec{k} b} \vec{\Pi}_{ \vec{k} a}  \right) \nonumber\\
&+&\frac{a^2(\eta)}{2} \vec{X}^T_{\vec{k}}\ \left(\Omega_k^2+\frac{\partial_{\eta}\left(a^2(\eta) {\mathcal M}_S\right)}{a^2(\eta)}+{\mathcal M}_A^T  {\mathcal O}^{-1} {\mathcal M}_A\right)\vec{X}_{-\vec{k}}.
\end{eqnarray}
Ultimately, since the differences in the various operator ordering possibilities are proportional to the trace of ${\mathcal O}^{-1}{\mathcal M}_A$ which vanishes identically, all orderings give the same result.

Quantization of this system now proceeds in the standard way, demanding that $\vec{\Pi}_{\vec{k}},\ \vec{X}_{\vec{q}}$ satisfy the commutation relations:
\begin{equation}
\label{eq:commutation}
\left[\vec{X}_{\vec{q}a}, \vec{\Pi}_{\vec{k}b}\right]= i\delta_{a b} \left(2\pi\right)^3 \delta^{(3)}\left(\vec{q}+\vec{k}\right).
\end{equation}
In the Schrödinger picture, the commutation relations are enforced by the choice:
\begin{equation}
\label{eq:commmomenta}
 \vec{\Pi}_{\vec{k}a}=-i \left(2\pi\right)^3 \frac{\delta}{\delta \vec{X}_{-\vec{k}a}}, \quad a,b=1,2,
\end{equation}
acting on wavefunctionals that depend on $\vec{X}_{\vec{q}a}$ and time.

Next we turn to the representation of entangled states in the Schrödinger picture and the form of the Schrödinger equation for them.

\subsection{The Functional Schrödinger Equation for Entangled Wavefunctionals}
\label{subsec:schroeq}

As discussed in section \ref{sec:intro}, our aim is to construct a class of states that entangle $\zeta$ with the field fluctuations $\chi$, yet remain Gaussian. Thus we write the wavefunctional $\Psi$ as:
\begin{equation}
\label{eq:entagledstate}
\Psi\left[\left\{\zeta_{\vec{k}}\right\}, \left\{\chi_{\vec{k}}\right\};\eta\right]=N(\eta) \exp\left(-\frac{1}{2}\int \frac{d^3 k}{(2\pi)^3}\ \vec{X}^T_{\vec{k}}\ {\mathcal K}_k(\eta)\ \vec{X}_{-\vec{k}}\right).
\end{equation}
Here, ${\mathcal K}_k(\eta)$ is a matrix of kernels:
\begin{equation}
\label{eq:kernels}
{\mathcal K}_k(\eta)=\left(\begin{array}{cc}A_k(\eta) & C_k(\eta) \\C_k(\eta) & B_k(\eta)\end{array}\right), 
\end{equation}
with $C_k(\eta)$ encoding the entanglement between the fluctuations. 

We now use this wavefunctional in the Schrödinger equation \eqref{eq:Seqn}. The strategy will be to compute both sides separately and then match the powers of $\vec{X}_{\vec{k}}$ that appear. The left hand side is given by:
\begin{equation}
\label{eq:SeqnLeft}
i\partial_{\eta} \Psi\left[\left\{\zeta_{\vec{k}}\right\}, \left\{\chi_{\vec{k}}\right\};\eta\right]=\left(i\frac{N'(\eta)}{N(\eta)}-\frac{i}{2}\langle \vec{X}^T_{\vec{k}}\ {\mathcal K}^{\prime}_k(\eta)\ \vec{X}_{-\vec{k}}\rangle\right)\Psi\left[\left\{\zeta_{\vec{k}}\right\}, \left\{\chi_{\vec{k}}\right\};\eta\right],
\end{equation}
where the angular brackets denote the $k$-space integral, including the factor of $(2\pi)^{-3}$. On the right hand side we note that the factors of $(2\pi)^{3}$ between the expression of the momentum as a derivative in \eqref{eq:commmomenta} and in the $k$-space measure cancel when the momenta act on the exponential. We can also simplify things in advance by noting that:
\begin{eqnarray}
\label{eq:mixingop}
\frac{1}{2}\left[{\mathcal O}^{-1}{\mathcal M}_A\right]_{a b}\left(\vec{\Pi}_{ \vec{k} a} \vec{X}_{ -\vec{k} b} + \vec{X}_{ -\vec{k} b} \vec{\Pi}_{ \vec{k} a}  \right) &=& \left[{\mathcal O}^{-1}{\mathcal M}_A\right]_{a b}\vec{X}_{ -\vec{k} b} \vec{\Pi}_{ \vec{k} a}\nonumber\\
&-& \frac{i (2\pi)^{3}}{2}\delta^{(3)}\left(\vec{q}=\vec{0}\right){\rm tr}\left({\mathcal O}^{-1}{\mathcal M}_A\right),
\end{eqnarray}
and we recognize $(2\pi)^{3}\delta^{(3)}\left(\vec{q}=\vec{0}\right)$ as the volume factor ${\mathcal V}$ that would appear in box quantization of the system. All terms containing this factor will contribute to the evolution of the normalization factor $N(\eta)$. But since ${\mathcal O}$ is symmetric while ${\mathcal M}_A$ is antisymmetric, the trace vanishes identically.

To compute the right hand side of \eqref{eq:Seqn}, we first compute the action of $\vec{\Pi}_{\vec{k}}$ on the wavefunctional:
\begin{equation}
\vec{\Pi}_{\vec{k} a}\Psi\left[\left\{\zeta_{\vec{k}}\right\}, \left\{\chi_{\vec{k}}\right\};\eta\right]=\left(i {\mathcal K}_k (\eta) \vec{X}_{\vec{k}}\right)_a \Psi\left[\left\{\zeta_{\vec{k}}\right\}, \left\{\chi_{\vec{k}}\right\};\eta\right].
\end{equation}
The second application of a momentum operator, as present in the kinetic term of the $k$-space integrated Hamiltonian, $\langle \vec{\Pi}^T_{\vec{k}}\ {\mathcal O}^{-1}\  \vec{\Pi}_{-\vec{k}}\rangle$, will bring down another factor of $\left(i {\mathcal K}_k (\eta) \vec{X}_{\vec{k}}\right)$, as well as a term proportional to the box volume ${\mathcal V}$:
\begin{eqnarray}
\label{eq:kineticaction}
&\frac{1}{2}\langle \vec{\Pi}^T_{\vec{k}}\ {\mathcal O}^{-1}\  \vec{\Pi}_{-\vec{k}}\rangle \Psi\left[\left\{\zeta_{\vec{k}}\right\}, \left\{\chi_{\vec{k}}\right\};\eta\right]=\nonumber\\
&\frac{1}{2}\left({\mathcal V}\langle {\rm tr}\left[{\mathcal O}^{-1} {\mathcal K}_k\right]\rangle-
 \langle \vec{X}^T_{\vec{k}} \left({\mathcal K}_k^T {\mathcal O}^{-1} {\mathcal K}_k\right) \vec{X}_{-\vec{k}}\rangle\right) \Psi\left[\left\{\zeta_{\vec{k}}\right\}, \left\{\chi_{\vec{k}}\right\};\eta\right]\end{eqnarray}

The next term to deal with is the operator in \eqref{eq:mixingop}:
\begin{equation}
\label{eq:mixingaction}
\langle \left[{\mathcal O}^{-1}{\mathcal M}_A\right]_{a b}\vec{X}_{ -\vec{k} b} \vec{\Pi}_{ \vec{k} a}\rangle  \Psi\left[\left\{\zeta_{\vec{k}}\right\}, \left\{\chi_{\vec{k}}\right\};\eta\right]=i \langle \vec{X}^T_{\vec{k}}  \left({\mathcal K}_k^T {\mathcal O}^{-1} {\mathcal M}_A\right) \vec{X}_{-\vec{k}} \rangle  \Psi\left[\left\{\zeta_{\vec{k}}\right\}, \left\{\chi_{\vec{k}}\right\};\eta\right].
\end{equation}
Combining \eqref{eq:kineticaction}, \eqref{eq:mixingaction} with the final term in \eqref{eq:Hamiltonian2} and matching powers of $ \vec{X}_{\vec{k}}$ gives the equations for the normalization and the kernel matrix:
\begin{subequations}\label{eq:eoms}
\begin{align}
\label{eq:eoms:1}
i\frac{N'(\eta)}{N(\eta)} & =\frac{1}{2}{\mathcal V}\langle {\rm tr}\left[{\mathcal O}^{-1} {\mathcal K}_k\right]\rangle
\\
\label{eq:eoms:2}
i {\mathcal K}_k^{\prime}(\eta) & = \frac{1}{a^2(\eta)}\left({\mathcal K}_k^T {\mathcal O}^{-1} {\mathcal K}_k\right)+i \left({\mathcal K}_k^T {\mathcal O}^{-1} {\mathcal M}_A+{\mathcal M}_A^T {\mathcal O}^{-1}  {\mathcal K}_k\right) \nonumber\\
& - {a^2(\eta)}\left(\Omega_k^2+\frac{\partial_{\eta}\left(a^2(\eta) {\mathcal M}_S\right)}{a^2(\eta)}+{\mathcal M} _A^T{\mathcal O}_k^{-1} {\mathcal M}_A\right),
\end{align}
\end{subequations}
where we symmetrized the middle term in order to be able to match independent powers of the modes.

We can decompose these equations into those for the individual kernels $A_k(\eta)$, $B_k(\eta)$, $C_k(\eta)$. Let's define:
\begin{equation}
\label{eq:kerneldefs}
z(\eta)=\sqrt{2 M_{\rm Pl}^2\ \epsilon\ a^2(\eta)}\ ,\ D=2 M_{\rm Pl}^2\ \epsilon-\left(\frac{\sigma^{\prime}}{{\mathcal H}}\right)^2.
\end{equation}
$D$ is a measure of how much of a spectator $\Sigma$ is, as it measures the relative sizes of the field contributions to the kinetic energy density. With these definitions, the equations of motion for the kernels are given by
\begin{subequations}\label{eq:kerneleqs}
\begin{align}
\label{eq:kerneleqs:A}
 i\partial_{\eta} A_k(\eta)  =& \left[- z(\eta)^2 k^2 +\frac{A_k^2}{ z(\eta)^2}\right] +\frac{1}{D}\left[ \left(\frac{\sigma^{\prime}}{{\mathcal H}}\right)\frac{A_k}{ z(\eta)}+\sqrt{2 M_{Pl}^2 \epsilon}\ \frac{C_k}{a(\eta)} + \right .\nonumber\\
& \left . \frac{i  z(\eta)}{2} \left((3-\epsilon) \sigma^{\prime}+\frac{a^2(\eta) \partial_{\sigma}V(\sigma)}{{\mathcal H}}\right)\right]^2,\\
\nonumber\\
\label{eq:kerneleqs:B}
i\partial_{\eta} B_k(\eta)  =& \left[-a^2(\eta)\left(k^2+a^2(\eta) \partial^2_{\sigma}V(\sigma)\right)+\frac{B_k^2}{a^2(\eta)}\right] \nonumber \\ &-a^2(\eta)\left[ (3-\epsilon) \frac{\sigma^{\prime 2}}{M_{\rm Pl}^2}+\frac{2 a^2(\eta)\ \sigma^{\prime}\ \partial_{\sigma}V(\sigma)}{{\mathcal H}\ M_{\rm Pl}^2}\right] \nonumber\\
&+ \frac{1}{D}\left[\frac{1}{a(\eta)}\left(C_k+\frac{\sigma^{\prime}}{{\mathcal H}} B_k\right)-\frac{i a(\eta)}{2}\left((3-\epsilon) \sigma^{\prime}+\frac{a^2(\eta) \partial_{\sigma}V(\sigma)}{{\mathcal H}}\right)\right]^2,\\
\nonumber\\
\label{eq:kerneleqs:C}
i\partial_{\eta} C_k(\eta) =& \frac{1}{a^2(\eta) D}\left[C_k\left(A_k+2 M_{\rm Pl}^2 \epsilon\ B_k\right)+\frac{\sigma^{\prime}}{{\mathcal H}}\left(C_k^2+A_k B_k\right)\right]\nonumber\\
&+\frac{1}{ D}\left[\left(-A_k+2 M_{\rm Pl}^2 \epsilon\ B_k\right)\left(\frac{i }{2} \left((3-\epsilon) \sigma^{\prime}+\frac{a^2(\eta) \partial_{\sigma}V(\sigma)}{{\mathcal H}}\right)\right)\right]\nonumber\\
& -\frac{a^2(\eta)}{D}\frac{\sigma^{\prime}}{{\mathcal H}}\left[\frac{i}{2}\left((3-\epsilon)\sigma^{\prime}+\frac{a^2(\eta) \partial_{\sigma}V(\sigma)}{{\mathcal H}}\right)\right]^2 \nonumber\\
&+ a^2(\eta) \left[\epsilon a^2(\eta) \partial_{\sigma}V(\sigma)+\frac{\sigma^{\prime}}{{\mathcal H}}\left(k^2+\frac{1}{2} a^2(\eta) \partial^2_{\sigma}V(\sigma)-\frac{1}{2}\ \epsilon\ \eta_{\rm sl} {\mathcal H}^2\right)\right].
\end{align}
\end{subequations}
Here $\eta_{\rm sl}$ denotes the second slow roll parameter $ \eta_{\rm sl}\equiv \epsilon'\slash {\mathcal H} \epsilon$. Note that, for completeness, we have kept the term $\eta_{\rm sl} \epsilon$, as well as terms quadratic in $\epsilon$ despite them being higher order in slow-roll. We will only keep the leading terms in slow-roll parameters when we turn to our numerical work. These equations should be solved in conjunction with the zero mode equation \eqref{eq:zeromodeclass}.

\subsection{Making Contact with the Bunch-Davies Modes}
\label{subsec:contact}
We will use the kernels $A_k, B_k, C_k$ in our further explorations of entanglement below. However, we recognize that this is a somewhat unorthodox way of constructing the power spectrum, as opposed to the standard way using the Bunch-Davies modes. Since we are calculating the same physical quantity, we expect that the kernels and the modes should be related. 

We have written equations \eqref{eq:kerneleqs:A}, \eqref{eq:kerneleqs:B} in a suggestive way. The first parentheses in each equation consists of the terms that would have been present when $\sigma$ is at the minimum of its potential. The remaining terms involve corrections that act as the sources of entanglement. Let's restrict our attention to the first set of terms and assume that $C=0,\ \sigma^{\prime}=\partial_{\sigma}V(\sigma)=0$. 

In this case equations \eqref{eq:kerneleqs:A}, \eqref{eq:kerneleqs:B} are decoupled Ricatti equations and there is a well-known prescription to convert the first order non-linear equation into a second order linear one. The most general Ricatti equation takes the form
\begin{equation}
\label{eq:Ricatti}
i K^{\prime}(\eta)=\alpha_2(\eta) K^2(\eta)+\alpha_1(\eta) K(\eta)+\alpha_0(\eta).
\end{equation}
Our goal is to transform this into a linear equation. Thus write
\begin{equation}
\label{eq:Ricatti_trans}
i K(\eta) = \frac{1}{\alpha_2(\eta)}\left(\frac{f'(\eta)}{f(\eta)}-\Delta(\eta)\right),
\end{equation}
where $\Delta$ is a term that allows us to at least partially control the final form of the second order equation. Inserting \eqref{eq:Ricatti_trans} into \eqref{eq:Ricatti}, we arrive at:
\begin{equation}
\label{eq:RicattiToMode}
f''(\eta)+\left(i\alpha_1-\frac{\alpha_2^{\prime}}{\alpha_2}-2\Delta\right)f'(\eta)+\left(-\alpha_0 \alpha_2-i \alpha_1\Delta+\Delta^2+\frac{\alpha_2^{\prime}}{\alpha_2}\Delta-\Delta^{\prime}\right)f(\eta)=0
\end{equation}
We see that we have the freedom to choose $\Delta$ to eliminate the term linear in $f'(\eta)$: $2\Delta=i\alpha_1-\alpha_2^{\prime}\slash \alpha_2$. Doing this leads to:
\begin{equation}
\label{eq:RicattiToModeFinal}
f''(\eta)+\Omega^2 f(\eta)=0,\ \Omega^2=\frac{1}{4}\alpha_1^2-\alpha_0 \alpha_2-\frac{i}{2} \alpha_1^{\prime}+\frac{i \alpha_1 \alpha_2^{\prime}}{2 \alpha_2}-\frac{3}{4} \left(\frac{\alpha_2^{\prime}}{\alpha_2}\right)^2+\frac{\alpha_2^{\prime \prime}}{2 \alpha_2}.
\end{equation}
Applying this procedure to equation~\eqref{eq:kerneleqs:A} yields
\begin{equation}
\label{eq:ZetaMode}
f''(\eta)+\left(k^2-\frac{z''(\eta)}{z(\eta)}\right)f(\eta)=0,
\end{equation}
which is exactly the mode equation that gives rise to the Bunch-Davies modes for $\zeta$~\cite{Baumann2011}. Likewise, applying the Ricatti procedure to \eqref{eq:kerneleqs:B} gives us:
\begin{equation}
\label{eq:MassMode}
g''(\eta)+\left(k^2+ a^2(\eta)\left . \partial_{\sigma}^2 V(\sigma) \right |_{\sigma=\sigma_{\rm min}}-\frac{a''(\eta)}{a(\eta)}\right)g(\eta)=0,
\end{equation}
where $\sigma_{\rm min}$ is the location of a minimum of $V(\sigma)$. This is again seen to be the mode equation expected for a massive field.

All three of our kernels satisfy Ricatti equations, so we could generate coupled mode equations as above and solve those. This is what was done in~\cite{Albrecht:2014aga,Bolis:2016vas,Bolis:2019fmq}, though in the situations discussed there, the entanglement kernel equation was not a Ricatti one, but was already linear. However, as shown in~\cite{Albrecht:2014aga}, the power spectrum is most directly accessible through the real parts of the kernels and as we are computing the power spectrum numerically, we may as well solve for the kernels directly via equation \eqref{eq:kerneleqs}.

We \emph{will} need to use the relation between kernels and modes when discussing the initial conditions for the kernels. We will choose to match the \emph{modes} to the standard Bunch-Davies ones at the initial time $\eta_0$ at which the entangled evolution begins. This allows us to compare the resulting power spectrum directly with the standard non-entangled case. The Ricatti relation then allows us to use the initial conditions for the modes to get at those for the kernels, although we are taking the initial entanglement to vanish, $C(\eta_0)=0$. For later reference, we write the real and imaginary parts of a generic kernel in terms of the modes:
\begin{subequations}\label{eq:realimKernels}
\begin{align}
\label{eq:realinKernels:real}
K_R =& \frac{i W\left[f,f^*\right]}{2 \alpha_2\left| f\right|^2}-\frac{\Delta_I}{\alpha_2}\\
\label{eq:realinKernels:imag}
K_I =& -\frac{1}{2 \alpha_2}\partial_{\eta}\ln\left | f\right |^2+\frac{\Delta_R}{\alpha_2},
\end{align}
\end{subequations}
where $W\left[f,f^*\right]$ is the Wronskian between the mode and its complex conjugate. This is constant in the non-entangled case, but will not remain so once entanglement is included. However, since we are making use of equations~\eqref{eq:realimKernels} only to help set initial conditions for the kernels, we can choose the \emph{initial} value of the Wronskian such that $i W\left[f,f^*\right](\eta_0)=1$. Using the relationship between $\Delta$ and the coefficient functions $\alpha_1, \alpha_2$, we can write
\begin{subequations}\label{eq:realimKernelsIC}
\begin{align}
\label{eq:realinKernelsIC:real}
K_R(\eta_0) =& \frac{1}{2 \alpha_2(\eta_0)}\left(\frac{1}{\left| f(\eta_0)\right|^2}-\alpha_{1 R}(\eta_0)\right)\\
\label{eq:realinKernelsIC:imag}
K_I(\eta_0) =& -\frac{1}{2 \alpha_2(\eta_0)}\left(\left . \partial_{\eta}\ln\left | f\right |^2\right |_{\eta=\eta_0}+\alpha_{1 I}(\eta_0)+\left . \frac{\alpha_2^{\prime}}{\alpha_2}\right |_{\eta=\eta_0}\right)\nonumber\\
= & -\frac{1}{2 \alpha_2(\eta_0)}\left(\left . \partial_{\eta}\ln \left(\alpha_2\ \left | f\right |^2\right)\right |_{\eta=\eta_0}+\alpha_{1 I}(\eta_0)\right),
\end{align}
\end{subequations}
where we have made use of the fact that for the $A_k$ and $B_k$ kernels, $\alpha_2$ is real and equal to $2 M_{\rm Pl}^2 \epsilon\slash z^2(\eta) D$ for $A_k$ and $2 M_{\rm Pl}^2 \epsilon\slash a^2(\eta) D$ for $B_k$. With the Wronskian condition above, the relevant modes we will use to match to the Bunch-Davies results at the initial time are:
\begin{subequations}\label{eq:initialmodes}
\begin{align}
\label{eq:initialmodes:zeta}
f_{\zeta}(\eta)= & \frac{\sqrt{-\pi \eta}}{2} H_{\nu_{\zeta}}^{(2)}(-k\eta),\ \nu_{\zeta}^2 = \frac{9}{4}+3\epsilon +\frac{3}{2}\eta_{\rm sl}\\
\label{eq:initialmodes:chi}
g_{\chi}(\eta)= & \frac{\sqrt{-\pi \eta}}{2} H_{\nu_{\chi}}^{(2)}(-k\eta),\ \nu_{\chi}^2 = \frac{9}{4}+3\epsilon -\frac{m^2}{(1-\epsilon)^2 H_{\rm dS}^2},
\end{align}
\end{subequations}
where $H_{\rm dS}$ is the Hubble parameter of the de Sitter spacetime occurring when $\epsilon=0$, and $H_{\rm dS}\equiv {\mathcal H}_0\slash a_0$, $a_0$ being the initial value of the scale factor and ${\mathcal H}_0$ that of the conformal time Hubble parameter. The mass parameter $m^2$ will be taken to be $\left|\partial^2_{\sigma}V(\sigma)\right|$ evaluated at the initial value of $\sigma$. This is equivalent to replacing the potential at the initial time by an upright mass term, which then switches to the full potential at $\eta_0$.

\section{The Entangled Power Spectrum and CMB Temperature Anisotropies}
\label{sec:powerspectrum}

We now turn to the main part of our project: to use the entangled state described above to compute the $\zeta$ power spectrum and the concomitant CMB temperature anisotropies. 

We start by using the results in~\cite{Albrecht:2014aga,Bolis:2016vas,Bolis:2019fmq} to write the $\zeta$ two-point function as:
\begin{equation}
\label{eq:zeta2pt}
\langle \zeta_{\vec{k}} \zeta_{\vec{k}^{\prime}} \rangle=(2\pi)^3 \delta^{(3)}\left(\vec{k}+\vec{k}^{\prime}\right)\left( \frac{B_{k R}}{2\left(A_{k R} B_{k R}-C_{k R}^2\right)}\right)\equiv (2\pi)^3 \delta^{(3)}\left(\vec{k}+\vec{k}^{\prime}\right) P_{\zeta}(k).
\end{equation}
The dimensionless form of the power spectrum~\cite{Baumann2011} is given by 
\begin{equation}
\label{eq:dimlessPS}
\Delta^2_s=\frac{k^3}{2\pi^2} P_{\zeta}(k).
\end{equation}
To see that this is in fact dimensionless note that, since $\zeta_{\vec{k}},\ \chi_{\vec{k}}$ have mass dimensions $-3$ and $-2$, respectively, the kernels have the following mass dimensions: $\left[A_k\right]= 3,\ \left[B_k\right]= 1,\ \left[C_k\right]= 2$. This means that $P_{\zeta}(k)$ has dimension $-3$ and thus $\Delta^2_s$ is indeed dimensionless. 

We next turn to the actual problem of solving equations \eqref{eq:kerneleqs}.
The first order of business in solving equations \eqref{eq:kerneleqs} numerically is to scale all the kernels and the zero mode to make them dimensionless. We also need to rescale the time variable $\eta$. The scale factor for a slow-roll spacetime is approximately given by 
\begin{equation}
\label{eq:slowrollscalefactor}
a(\eta)=a_0\left(-\frac{1}{(1-\epsilon) {\mathcal H}_0 \eta}\right)^{\frac{1}{1-\epsilon}}=a_0\left(\frac{\eta_0}{\eta}\right)^{\frac{1}{1-\epsilon}},
\end{equation}
where $\eta_0$ is to related to ${\mathcal H}_0$ via $(1-\epsilon) {\mathcal H}_0\eta_0=-1$. Physically, we can choose to have $\eta_0$ vary from being the time at which the largest length scale appearing on the CMB sky leaves the horizon to being a time in which shorter scales or higher wave numbers leave the horizon. The main physical constraint is that of having the energy density due to the difference between the entangled state and the Bunch-Davies state be small enough to allow for a sufficient number of e-folds to occur. Beyond this, we would treat $\eta_0$ as part of the set of parameters one would estimate. When considering the sample situations discussed below, we will let $\eta_0$ vary so as to exhibit the changes that would occur in the different measured power spectra. 

We scale the time and wavenumbers as:
\begin{equation}
\label{eq:dimlesstimewave}
\tau = -\frac{\eta}{\eta_0},\ q= \frac{k}{k_0}=\frac{k}{{\mathcal H}_0}=-(1-\epsilon) k\eta_0.
\end{equation}
Note that, since the conformal times are all negative, $\tau$ runs from $-1$ to $0$.

The scalings of $A_k$ and $B_k$ are essentially dictated by the parts of equations \eqref{eq:kerneleqs:A}, \eqref{eq:kerneleqs:B} in the first set of brackets, i.e., those that would have been present even in the absence of entanglement. We define dimensionless kernels $\tilde{A}_q,\ \tilde{B}_q$ as:
\begin{subequations}\label{eq:dimlessABkernels}
\begin{align}
\label{eq:dimlessABkernels:A}
A_k(\eta) =& \frac{ z(\eta)^2}{\left(-\eta_0\right)} \tilde{A}_q(\tau)\\
\label{eq:dimlessABkernels:B}
B_k(\eta) =& \frac{a^2(\eta)}{\left(-\eta_0\right)} \tilde{B}_q(\tau),
\end{align}
where $ z(\eta)$ is defined in equation \eqref{eq:kerneldefs}. Given the dimensions of $A_k$ and $B_k$, we see that $\tilde{A}_q,\ \tilde{B}_q$ are indeed dimensionless. For $C_k$ we use a combination of $ z(\eta)$ and the scale factor,
\begin{equation}
\label{eq:dimlessABkernels:C}
C_k(\eta) =  \frac{ z(\eta) a(\eta)}{\left(-\eta_0\right)} \tilde{C}_q(\tau).
\end{equation}
\end{subequations}

In terms of the dimensionless kernels and wavenumbers, the dimensionless power spectrum in \eqref{eq:dimlessPS} is given by:
\begin{equation}
\label{eq:dimlessKdimlessk}
\Delta^2_s =  A_s \left(q^3 (-\tau)^{2 \nu_{\zeta}-1}\right)\left(\frac{ \tilde{B}_{q R}}{\tilde{A}_{q R} \tilde{B}_{q R}-\tilde{C}_{q R}^2}\right),
\end{equation}
where
\begin{equation}
 A_s = \left(\frac{H_{dS}^2}{8\pi^2 \epsilon (1-\epsilon) M_{\rm Pl}^2}\right). 
\end{equation}
We have also used the definitions of $q$ and $\tau$ above as well as that of $z$ in terms of the scale factor, $\epsilon$ and $M_{\rm Pl}$, in addition to the relation $(1-\epsilon) {\mathcal H}_0\eta_0=-1$. We see that the dimensionless time $\tau$ makes an explicit appearance here. Since we are \emph{not} assured that the modes become frozen after horizon crossing, we will evaluate $\Delta^2_s $ in the late time limit $\tau\rightarrow 0^{-}$. As discussed in~\cite{Kinney:2005vj}, the horizon crossing approximation requires certain conditions to be met and these do not obtain in our situation.

The zero mode should be made dimensionless as well. We choose to use the Planck mass to scale $\sigma$ with: $\sigma(\eta)=M_{\rm Pl} s(\tau)$. We also construct a dimensionless version of the potential: $V(\sigma) =\Lambda^4 \bar{V}(s)$, where $\Lambda$ is the natural energy scale associated with $V(\sigma)$. Each derivative of $V(\sigma)$ with respect to $\sigma$ corresponds to a derivative of $\bar{V}(s)$ with respect to $s$ with a factor of $M_{\rm Pl}$ appropriately inserted.

The final task to accomplish is to now rewrite the kernel equations \eqref{eq:kerneleqs} in terms of dimensionless quantities: 
\begin{subequations}\label{eq:dimlesskerneleqs}
\begin{align}
\label{eq:dimlesskerneleqs:A}
 i\partial_{\tau} \tilde{A}_q(\tau)  =& \frac{i(2 +\eta_{\rm sl})}{\tau(1-\epsilon)} \tilde{A}_q+\left[-\left(\frac{q}{1-\epsilon}\right)^2 +\tilde{A}_q^2\right] \\
 &+ \frac{1}{\bar{D}}\Bigg[(1-\epsilon)(-\tau \partial_{\tau} s) \tilde{A}_q+\sqrt{2 \epsilon}\ \tilde{C}_q + \frac{i}{2}\left((3-\epsilon) \partial_{\tau}s+\frac{\mu^2}{1-\epsilon}\left(-\frac{\partial_s\bar{V}(s)}{\tau}\right)\right)\Bigg]^2, \nonumber \\
\nonumber \\
\label{eq:dimlesskerneleqs:B}
i\partial_{\tau} \tilde{B}_q(\tau)  =& \frac{2 i}{\tau(1-\epsilon)}\tilde{B}_q+\left[-\left(\left(\frac{q}{1-\epsilon}\right)^2+\frac{\mu^2\ \partial_s^2 \bar{V}(s)}{\tau^2(1-\epsilon)^2}\right)+\tilde{B}_q^2\right] \nonumber \\
&-\left[ (3-\epsilon) \left(\partial_{\tau} s\right)^2-\frac{2 \mu^2\ \partial_{\tau} s}{\tau (1-\epsilon)} \partial_s\bar{V}(s)\right] \\
& + \frac{1}{\bar{D}} \left[(1-\epsilon)(-\tau \partial_{\tau} s) \tilde{B}_q+\sqrt{2 \epsilon}\ \tilde{C}_q -\frac{i }{2}\left((3-\epsilon) \partial_{\tau}s+\frac{\mu^2}{1-\epsilon}\left(-\frac{\partial_s\bar{V}(s)}{\tau}\right)\right)\right]^2, \nonumber\\
\nonumber \displaybreak\\
\label{eq:dimlesskerneleqs:C}
i\partial_{\tau} \tilde{C}_q(\tau) =&\frac{i(4+\eta_{\rm sl})}{2\tau (1-\epsilon)} \tilde{C}_q+\frac{\sqrt{2\epsilon}}{\bar{D}}\left[\sqrt{2\epsilon}\ \tilde{C}_q \left(\tilde{A}_q+\tilde{B}_q\right)+(1-\epsilon)(-\tau\partial_{\tau}s)\left(\tilde{C}_q^2+\tilde{A}_q \tilde{B}_q\right)\right] \nonumber\\
&+\frac{\sqrt{2\epsilon}}{\bar{D}}\left[\left(-\tilde{A}_q+\tilde{B}_q\right)\left(\frac{i}{2}\left((3-\epsilon) \partial_{\tau}s+\frac{\mu^2}{1-\epsilon}\left(-\frac{\partial_s\bar{V}(s)}{\tau}\right)\right)\right)\right] \\
& -\frac{1}{\bar{D}\sqrt{2\epsilon}}(1-\epsilon)(-\tau\partial_{\tau} s)\left[\frac{i}{2}\left((3-\epsilon)\partial_{\tau}s+\frac{\mu^2}{1-\epsilon}\left(-\frac{\partial_s\bar{V}(s)}{\tau}\right)\right)\right]^2\nonumber\\
& + \left[\sqrt{\frac{\epsilon}{2}}\frac{\mu^2}{(1-\epsilon)^2 \tau^2}\partial_s\bar{V}(s) + \frac{(1-\epsilon)}{\sqrt{2\epsilon}}(-\tau\partial_{\tau}s)\left(\left(\frac{q}{1-\epsilon}\right)^2+\frac{\mu^2\ \partial_s^2 \bar{V}(s)-\epsilon\ \eta_{\rm sl}}{2\tau^2 (1-\epsilon)^2}\right)\right], \nonumber\\ \nonumber
\end{align}
\end{subequations}
where $\bar{D}= 2\epsilon-(1-\epsilon)^2(-\tau\partial_{\tau} s)^2$, and $\mu^2 = \Lambda^4\slash(M_{\rm Pl}^2 H_{dS}^2)$. 

The zero mode equation also needs to be made dimensionless, but this is easily done:
\begin{equation}
\label{eq:zmdimless}
\partial_{\tau}^2 s-\frac{2}{\tau(1-\epsilon)} \partial_{\tau} s+\frac{\mu^2}{\tau^2 (1-\epsilon)^2} \partial_s\bar{V}(s)=0.
\end{equation}

We turn next to the initial conditions for the kernels. While we have already discussed this above, there are a few points worth focusing on. Our aim is to be able to compare our results with the standard inflationary ones, i.e., for similar parameters $\epsilon, \eta_{\rm sl}$, we want to extract the effect of non-trivial entanglement relative to the no-entanglement case. Thus, we'll choose the entanglement kernel to vanish initially, so that it is generated only by dint of the behavior of the zero mode. Thus $C_k(\eta_0)=0$, or in terms of the dimensionless quantities above, $\tilde{C}_q(\tau=-1)=0$. The initial conditions of the zero mode are taken to be free parameters: $s(\tau=-1)=s_0,\ \left . \partial_{\tau}s\right|_{\tau=-1}=v_0$. 

For the $A_k, B_k$ kernels we follow the discussion leading to equations \eqref{eq:realimKernelsIC}. We can either view the entangled inflationary phase as arising during the last $55-60$ e-folds of inflation, so that the initial conditions are set by the prior phase of non-entangled inflation, or inflation only lasts the minimal amount needed in order to solve the horizon and flatness problems and we choose the initial conditions to be as to close to the non-entangled case as possible. Either way, we arrive at equations \eqref{eq:realimKernelsIC} with 
\begin{equation}
\label{eq:AkernelIC}
\alpha_{1 A}(\eta_0)=\frac{i}{D_0}\left(\frac{\sigma^{\prime}(\eta_0)}{{\mathcal H}_0}\right)\left((3-\epsilon)\sigma^{\prime}(\eta_0)+\frac{a_0^2 V^{\prime}(\sigma_0)}{{\mathcal H}_0}\right),\ \alpha_{2 A}(\eta_0)=\frac{1}{a_0^2 D_0},
\end{equation}
for the $A_k$ kernel and $\alpha_{1 B}(\eta_0)=-\alpha_{1 A}(\eta_0),\ \alpha_{2 B}(\eta_0)=\left(2 M_{\rm PL}^2 \epsilon\right) \alpha_{2 A}(\eta_0)$ for the $B_k$ kernel. Here $D_0$ is the initial value of the quantity $D$ defined in equation \eqref{eq:kerneldefs} and we have made use of the vanishing of the initial entanglement. From this, equations \eqref{eq:realimKernelsIC}, and the definitions of the dimensionless kernels equations \eqref{eq:dimlesskerneleqs}, we can write the initial conditions for the real and imaginary parts of the $\tilde{A}_q, \tilde{B}_q$ kernels:
\begin{subequations}\label{eq:AkernelICs}
\begin{align}
\label{eq:AkernelICs:real}
 \tilde{A}_{q R}(\tau=-1)= &\left(\frac{\bar{D}_0}{2\epsilon}\right)\left(\frac{2}{\pi \left |H^{(2)}_{\nu_{\zeta}}(\frac{q}{1-\epsilon})\right|^2}\right)\\
\label{eq:AkernelICs:imag}
\tilde{A}_{q I}(\tau=-1)= &\frac{\bar{D}_0}{4\epsilon (1-\epsilon)}\left(3-\epsilon+q \left . \partial_x \ln \left|H^{(2)}_{\nu_\zeta}(x)\right|^2\right|_{x=\frac{q}{1-\epsilon}}\right)+\frac{\eta_{\rm sl}}{2(1-\epsilon)}+\nonumber\\
&+\frac{\eta_{\rm sl} v_0^2}{2}+ \frac{v_0}{4\epsilon}\left((1-\epsilon)(3-\epsilon) v_0+\mu^2 \left . \partial_s \bar{V}(s)\right|_{s=s_0}\right),
\end{align}
\end{subequations}
where we recall $\left . \partial_{\tau}s\right|_{\tau=-1}=v_0$ and $\bar{D}_0=2\epsilon-(1-\epsilon)^2 v_0^2$. An interesting point to note concerns the factors of $\epsilon$ in the denominators of both equations \eqref{eq:dimlesskerneleqs} and equations \eqref{eq:AkernelICs}. These seem worrisome in the $\epsilon\rightarrow 0$ limit; however in this case $\zeta$ is a pure gauge degree of freedom so our analysis is moot. 

A similar analysis for the $\tilde{B}_q$ kernel yields
\begin{subequations}\label{eq:BkernelICs}
\begin{align}
\label{eq:BkernelICs:real}
 \tilde{B}_{q R}(\tau=-1)= &\left(\frac{\bar{D}_0}{2\epsilon}\right)\left(\frac{2}{\pi \left |H^{(2)}_{\nu_{\chi}}(\frac{q}{1-\epsilon})\right|^2}\right)\\
\label{eq:BkernelICs:imag}
\tilde{B}_{q I}(\tau=-1) = &\frac{\bar{D}_0}{4\epsilon (1-\epsilon)}\left(3-\epsilon+q \left . \partial_x \ln \left|H^{(2)}_{\nu_\chi}(x)\right|^2\right|_{x=\frac{q}{1-\epsilon}}\right)+\frac{(1+\epsilon) v_0^2 }{4\epsilon}\eta_{\rm sl} + \nonumber\\
&+ \frac{3 v_0}{4\epsilon}\left((1-\epsilon)(3-\epsilon) v_0+\mu^2 \left . \partial_s \bar{V}(s)\right|_{s=s_0}\right).
\end{align}
\end{subequations}

Before turning to the numerical solution to these equations, we enumerate the various constraints we need to satisfy in order that we may treat the field $\Sigma$ as a spectator field. First, its energy density should be much less than that of the inflaton:
\begin{subequations}\label{eq:spectatorconstraints}
\begin{equation}
\label{eq:spectatorconstraints_potential}
V(\sigma)\ll M_{\rm Pl}^2 H_{\rm dS}^2,
\end{equation}
and second, the kinetic energy of the inflaton, encoded in the quantity $z(\eta)$ above should be larger than that of the spectator
\begin{equation}
\label{eq:spectatorconstraints_kinetic}
 \left(\frac{\sigma^{\prime}}{{\mathcal H}}\right)^2\ll2 M_{\rm Pl}^2 \epsilon.
\end{equation}
\end{subequations}
In terms of the dimensionless quantities introduced earlier, these constraints become:
\begin{subequations}\label{eq:dimlessspectatorconstraints}
\begin{align}
\label{eq:dimlessspectatorconstraints_potential}
\bar{V}(s)\ll \frac{1}{\mu^2},\\
\label{eq:dimlessspectatorconstraints_kinetic}
\left | -\tau \partial_{\tau}s\right |\ll \frac{\sqrt{2 \epsilon}}{1-\epsilon}.
\end{align}
\end{subequations}

There is another constraint we have to satisfy. We require that the wavefunctional be normalizable. For each $\vec{k}$ we demand
\begin{equation}
\label{eq:wfnormalization}
\int \Pi_{\vec{k}}\ {\mathcal D}^2\zeta_{\vec{k}}\ {\mathcal D}^2\chi_{\vec{k}}\  \left| \Psi\left[\left\{\zeta_{\vec{k}}\right\},\left\{\chi_{\vec{k}}\right\};\eta\right]\right|^2<\infty,
\end{equation}
where the wavefunctional is given in equation \eqref{eq:entagledstate}.
This requirement is equivalent to that of demanding that the two eigenvalues of
\begin{equation}
\label{eq:realquadform}
{\mathcal K}_{k R}(\eta)  = \begin{pmatrix}
	A_{k R}(\eta) &\ C_{k R}(\eta)\\
	 C_{k R}(\eta) &\ B_{k R}(\eta)
	\end{pmatrix},
\end{equation}
be positive. This in turn is equivalent to demanding both that $A_{k R}(\eta)+B_{k R}(\eta)$ be positive, as well as that the determinant $A_{k R}(\eta)B_{k R}(\eta)-C_{k R}^2(\eta)$ also be positive. The first constraint holds automatically as can be seen using the Ricatti trick as in equations \eqref{eq:realimKernels}. In our case $\alpha_{1 R}(\eta)=0$ for both the $A$ and $B$ kernels and $\alpha_2(\eta)$ is positive as long as equations \eqref{eq:spectatorconstraints} hold, so that both $A_{k R}(\eta)$ and $B_{k R}(\eta)$ are proportional to the modulus squared of a mode, with positive proportionality constants. The determinant constraint has to be checked during the time evolution. 

We will consider two different potentials for the spectator: a mass term and an axion-type periodic potential. For each case, we will choose some parameters that help display interesting features of the power spectrum as well as the TT and TE CMB anisotropies. In this work, we will \emph{not} perform an exhaustive parameter search, deferring that to later work.

We turn to the task of obtaining the power spectrum as given in equation \eqref{eq:dimlessKdimlessk} and from thence obtaining the various CMB anisotropies power spectra. For this last step, we use the CLASS Boltzmann solver~\cite{Blas_2011}. 

The standard, non-entangled, scalar inflationary power spectrum is typically written as\footnote{We have adjusted notation slightly from equation (38a) in~\cite{2020Planck} here, to make a smoother comparison with our equations.}:
\begin{equation}
    \Delta^{2}_{s}(k) = A_{s,k} \left( \frac{k}{k_{\rm{piv}}} \right)^{n(k)},
\end{equation}
where $k_{\rm{piv}}$ is the pivot scale and $n(k) = n_{s} - 1$ if one ignores the running of the spectral index $n_{s}$ \cite{2020Planck}. 

When our code generates the dimensionless power spectrum, $\Delta_{s}^{2} (q)$, (see equation \eqref{eq:dimlessKdimlessk}), we actually calculate and plot it in units of $A_{s}$, so we are in actuality plotting a parameter
\begin{equation}
   \tilde{\Delta}_{s}^{2} (q) = \left(q^3 (-\tau)^{2 \nu_{\zeta}-1}\right)\left(\frac{ \tilde{B}_{q R}}{\tilde{A}_{q R} \tilde{B}_{q R}-\tilde{C}_{q R}^2}\right) .
\end{equation}
To compare this with the latest Planck CMB data release~\cite{2020Planck}, we need to find a parameter, $f$, such that the following is true for the non-entangled case of our equations:
\begin{equation}
\label{f_convert}
    f  \tilde{\Delta}_{s, \rm NE}^{2} (q) =  \Delta^{2}_{s}(k) = A_{s,k} \left( \frac{k_{0} q}{k_{\rm{piv}}} \right)^{n_{s} - 1},
\end{equation}
where we used the conversion $k = q k_{0}$ from equation \eqref{eq:dimlesstimewave} and substituted $n(k)$ for $n_{s} -1$ for simplicity. Using Planck values~\cite{2020Planck} for $A_{s,k}$, $n_{s}$, and $k_{\rm{piv}}$, we can determine $f$ and then employ it to rescale our data. 

For both the non-entangled and entangled cases, where $k_{0} = 10^{-6}\ {\rm Mpc}^{-1}$ is chosen to be the largest observable scale, the parameter $f$ described above allows us to rescale $\tilde{\Delta}_{s}^{2} (q)$. Since the entangled power spectra are in essence corrections around a non-entangled baseline, it makes sense to use the non-entangled value to rescale them as well. We then input the resulting $\Delta^{2}_{s}(k)$ into CLASS to generate the $C_{l}$ plots in the subsequent sections. For the cases where we shift the onset of entanglement---which corresponds to shifting $k_{0}$ as discussed in section \ref{subsec:k0shift}---we obtain a slightly different $f$ in equation \eqref{f_convert}, but otherwise the data processing is exactly the same as the non-shifted case.

\subsection{Free Massive Scalar}
\label{subsec:fmassscalar}

We start by considering the case of a free massive spectator scalar: $V(\sigma)=m^2 \sigma^2\slash 2$. In this case, $\Lambda^4 = M_{\rm Pl}^2 m^2$, $\mu^2 = m^2\slash H_{\rm dS}^2$ and $\bar{V}(s) =s^2\slash 2$. 

\begin{figure}[hbtp]
    \centering
    \begin{subfigure}[b]{0.59\linewidth}
        \centering
        \includegraphics[width=\linewidth]{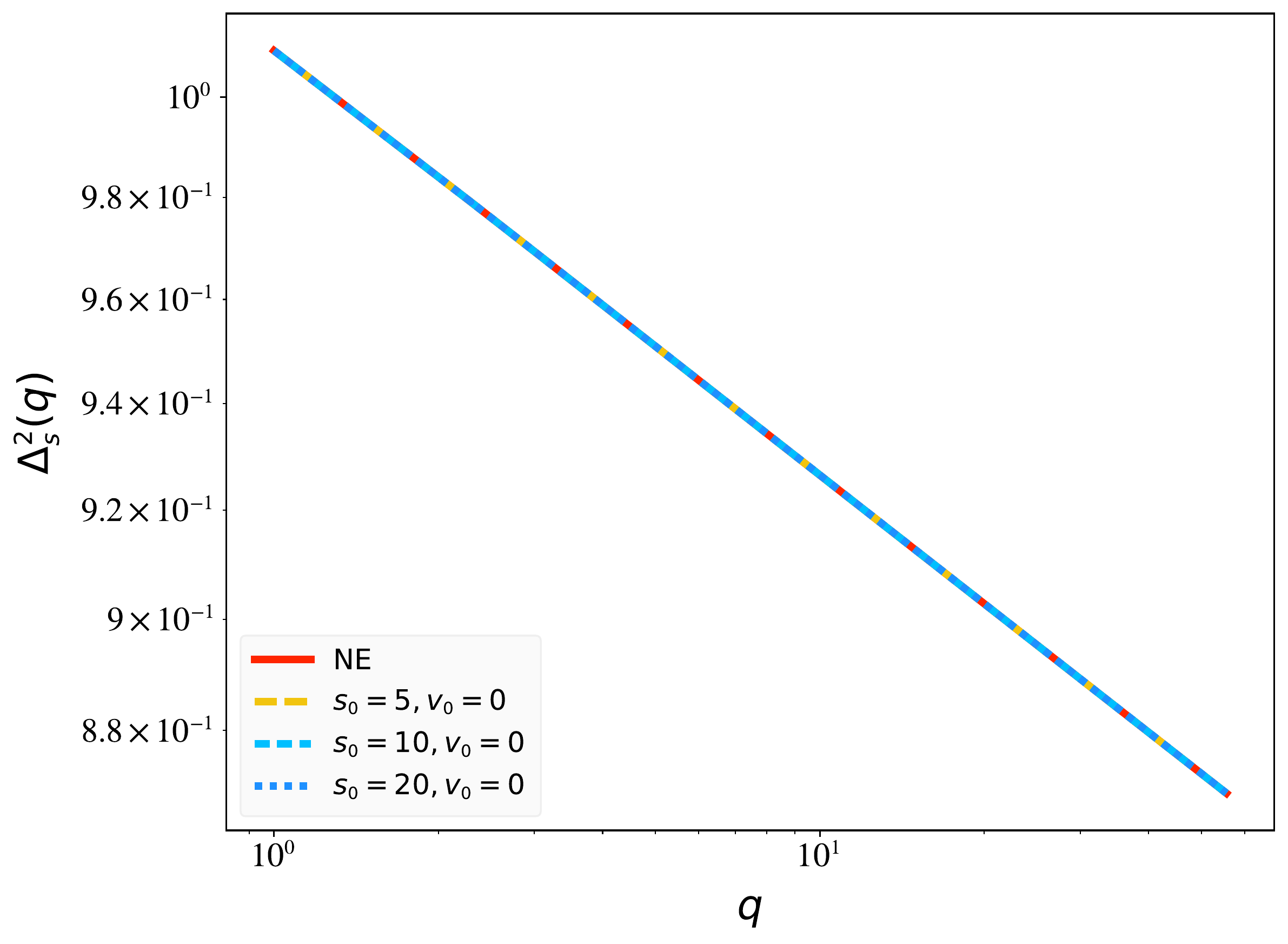}
        \subcaption{$\mu=0.01,\ v_0=0$}
    \end{subfigure}
    \begin{subfigure}[b]{0.59\linewidth}
        \centering
        \includegraphics[width=\linewidth]{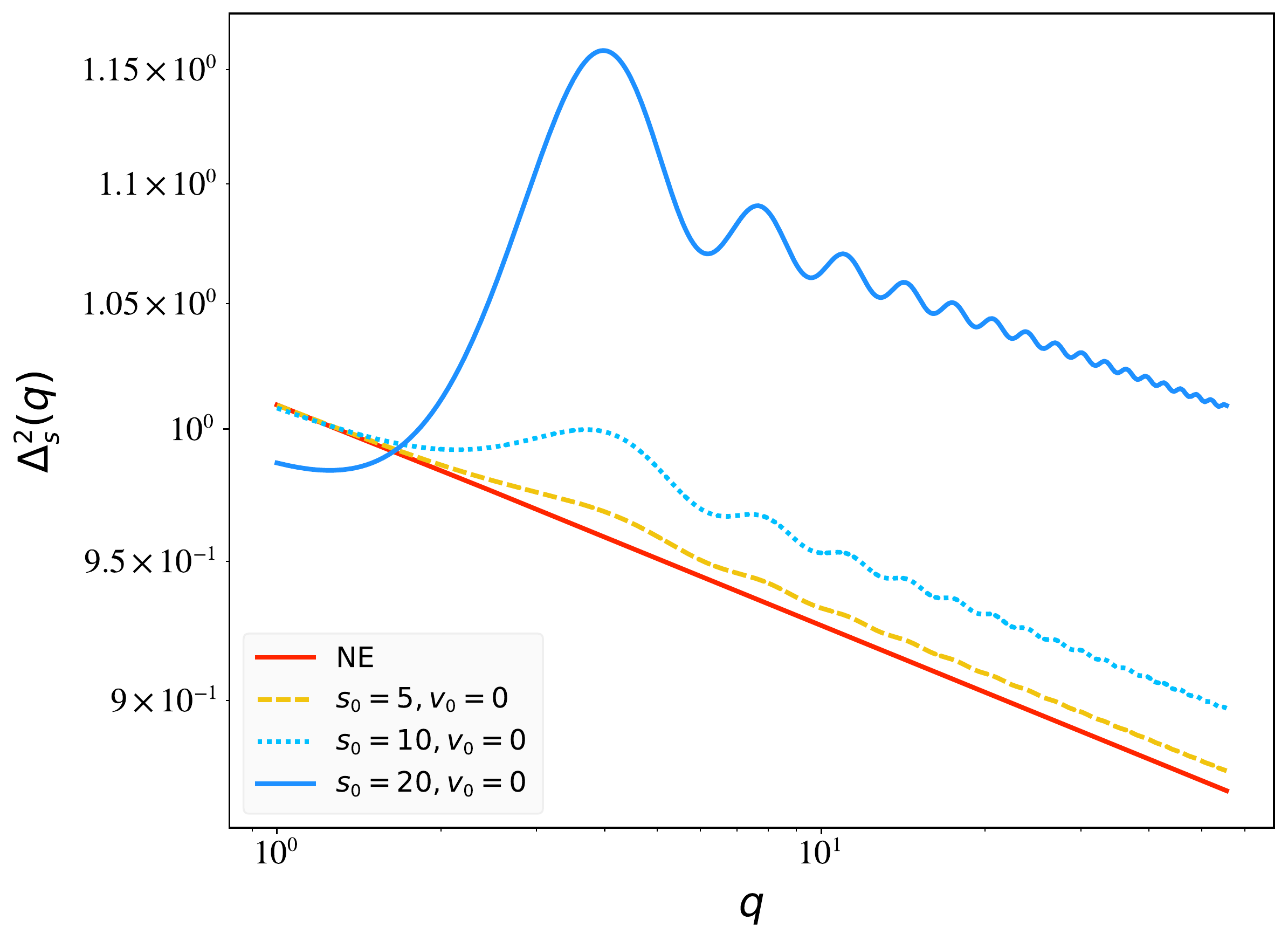}
        \subcaption{$\mu=0.1,\ v_0=0$}
    \end{subfigure}
    \begin{subfigure}[b]{0.59\linewidth}
        \centering
        \includegraphics[width=\linewidth]{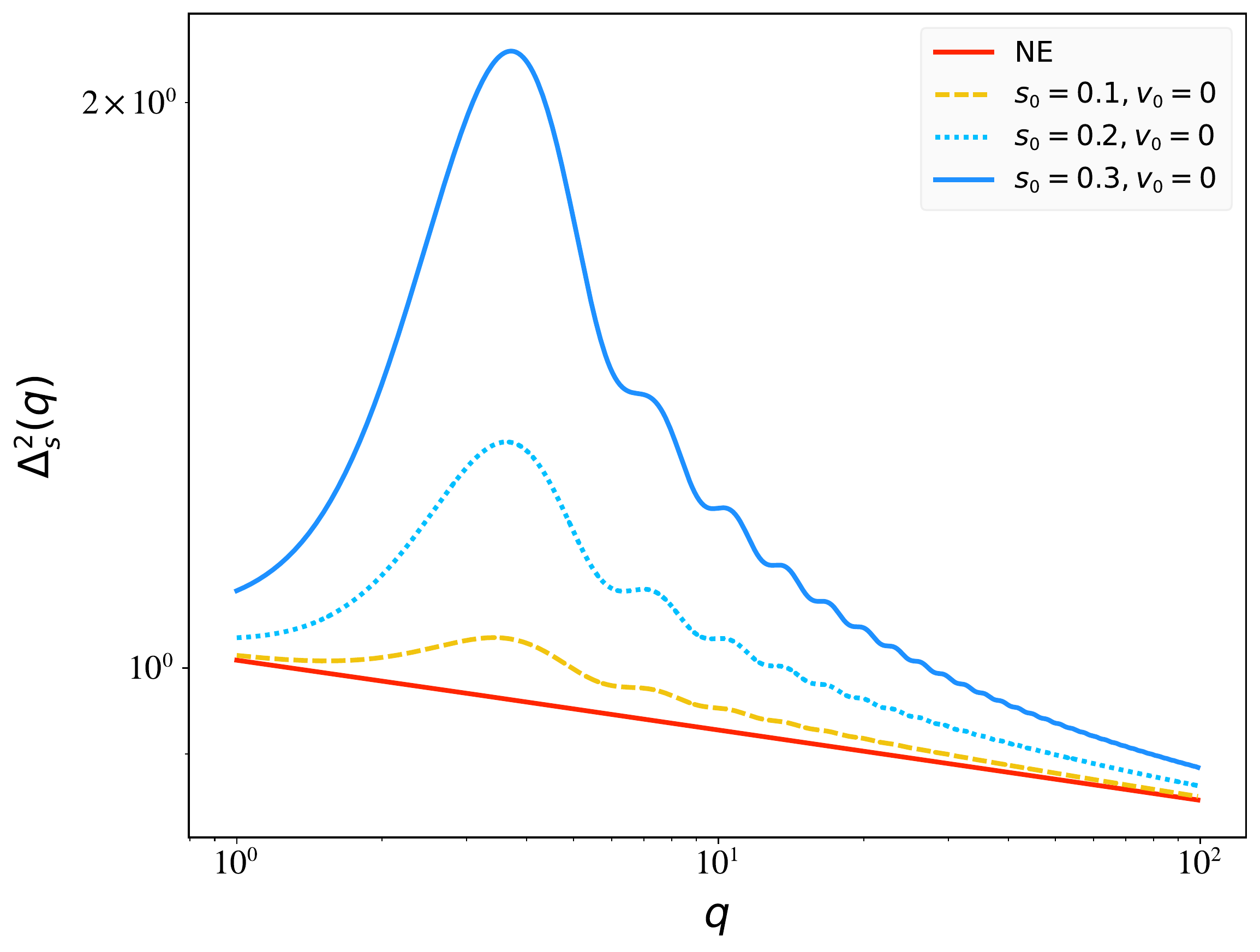}
        \subcaption{$\mu=1,\ v_0=0$}
    \end{subfigure}
    \caption{Log-log plots of the power spectrum $\Delta_s^2$ plotted in units of $A_s$ as a function of $q=k\slash {\mathcal H}_0$, for different values of $\mu$ and $s_0$. In all cases, $v_0$ is taken to be $0$ while $\mu = 0.01, 0.1,$ or $1$ in (a), (b), and (c), respectively. }
  \label{fig:zerovelocitycase}
\end{figure}

In Figure \ref{fig:zerovelocitycase} we see that, for low enough $\mu$ and $v_0=0$, the power spectra for the differing values of the initial field do not vary significantly from the non-entangled case ($s_0=0,v_0=0$). However, as soon as $\mu\sim 0.1$, features manifest themselves. In particular, damped oscillatory behavior is superposed over the non-entangled power spectrum; there is also an enhancement in the power after $q\sim 2$. For particularly high values of $\mu$ the oscillations are not only damped but, after an initial enhancement of power, the entangled power spectrum decays to match the non-entangled one for higher $q$ values (see also Figure \ref{{fig:ClandPkmu1nov0}}).

What happens as we allow for non-zero $v_0$?
\begin{figure}[hbtp]
\centering
  \includegraphics[scale=0.38]{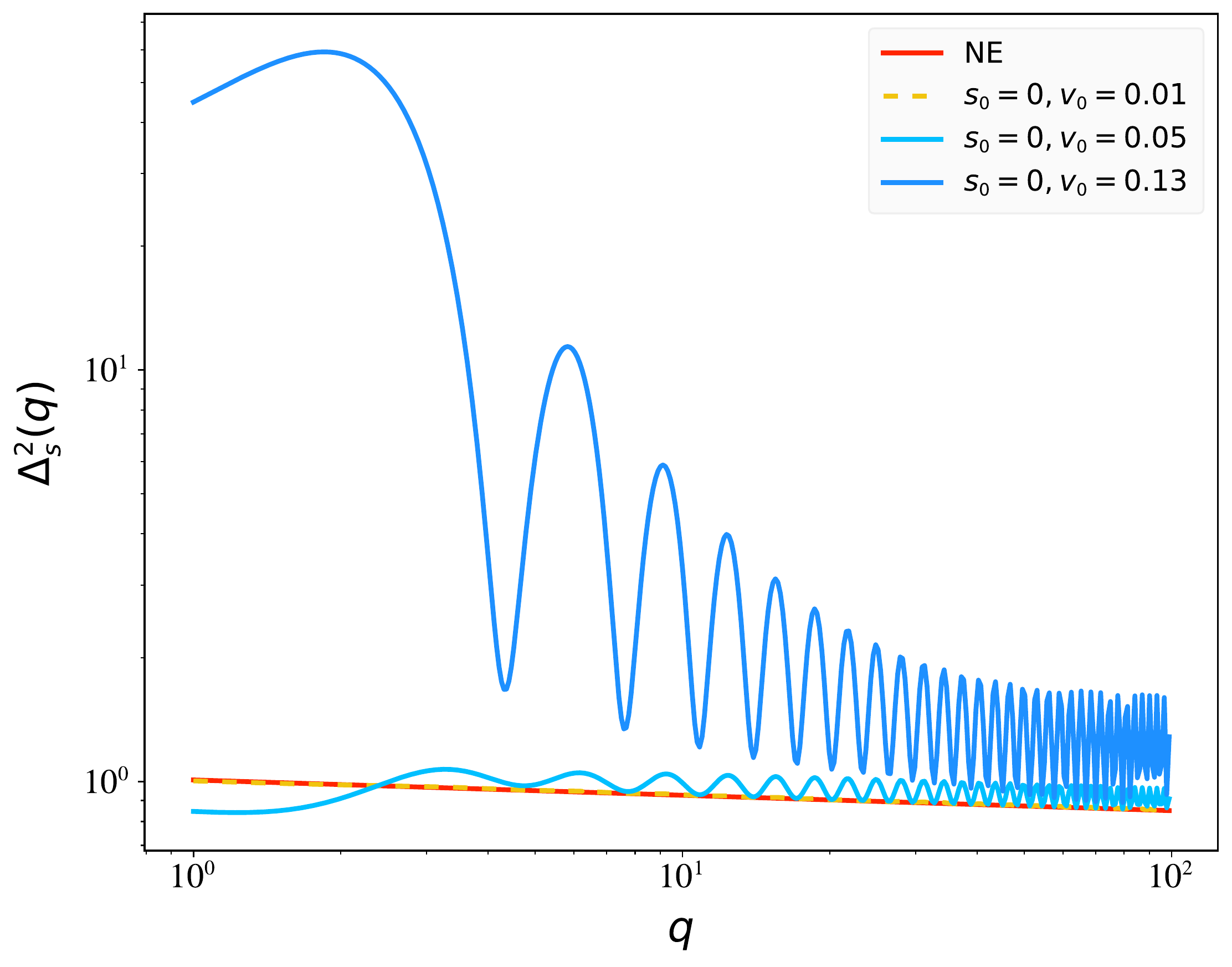}
  \vspace{-0.2cm}
\caption{Log-log plot of the power spectrum $\Delta_s^2$ plotted in units of $A_s$ as a function of $q=k\slash {\mathcal H}_0$, for $s_0=0$, $\mu=0.01$, and different choices of $v_0$.}
\label{{fig:zeropositioncase}}
\end{figure}
In Figure \ref{{fig:zeropositioncase}}, we see that we once again generate oscillations in the power spectrum. The value $v_0=0.13$ is near the boundary delineated in equation \eqref{eq:dimlessspectatorconstraints_kinetic}. For this case, the damped regime of the oscillations lasts for a couple of decades in $q$. But for the value $v_0=0.05$ the damped regime is shorter. In both cases, the oscillations stabilize at large enough $q$ and the troughs of the oscillations eventually sit directly on top of the non-entangled power spectrum, (see also Figure \ref{{fig:ClandPkmu0.1nos0}} for the case $v_0 = 0.05$), though it takes the higher $v_0$ value longer to exhibit this behavior. 

We can use these power spectra as initial conditions for a Boltzmann solver to the various CMB power spectra. In this paper we used CLASS~\cite{Blas_2011} to generate the TT and TE power spectra, given our dimensionless power spectra data and the data processing described in section~\ref{sec:powerspectrum}. For the TT and TE graphs generated in CLASS, we look at the unlensed power spectra, with input values for $h_{0}$, $\Omega_{b}$ and other required parameters taken from the unlensed values in Table 2 of~\cite{2020Planck}. We then compare our results with Planck data in the resulting graphs. For this portion of our analysis, we chose a representative subset of the initial parameters that generate the primordial power spectra in Figures \ref{fig:zerovelocitycase} - \ref{{fig:zeropositioncase}} to investigate the range of possible effects on the CMB power spectra. 

Figure \ref{{fig:ClandPkmu0.01nov0}} shows the primordial spectra in $k$ for the standard non-entangled case and the entangled case with $\mu=0.01$, $s_0=10$, and $v_0 =0$. The corresponding TT and TE angular power spectra are also shown and compared with data from Planck~\cite{2020Planck}. Predictably, since the primordial power spectra for the entangled and non-entangled case are identical by eye for these parameters, the TT and TE spectra are also indistinguishable. For this set of parameters, one could argue it is hard to tell whether Planck is ``seeing'' evidence of the BD state or evidence of a state of entanglement with a low mass scalar, since both scenarios appear observationally indistinguishable.
\begin{figure}[!hbtp]
\centering
  \begin{subfigure}[b]{0.62\linewidth}
  \centering
    \includegraphics[width=\linewidth]{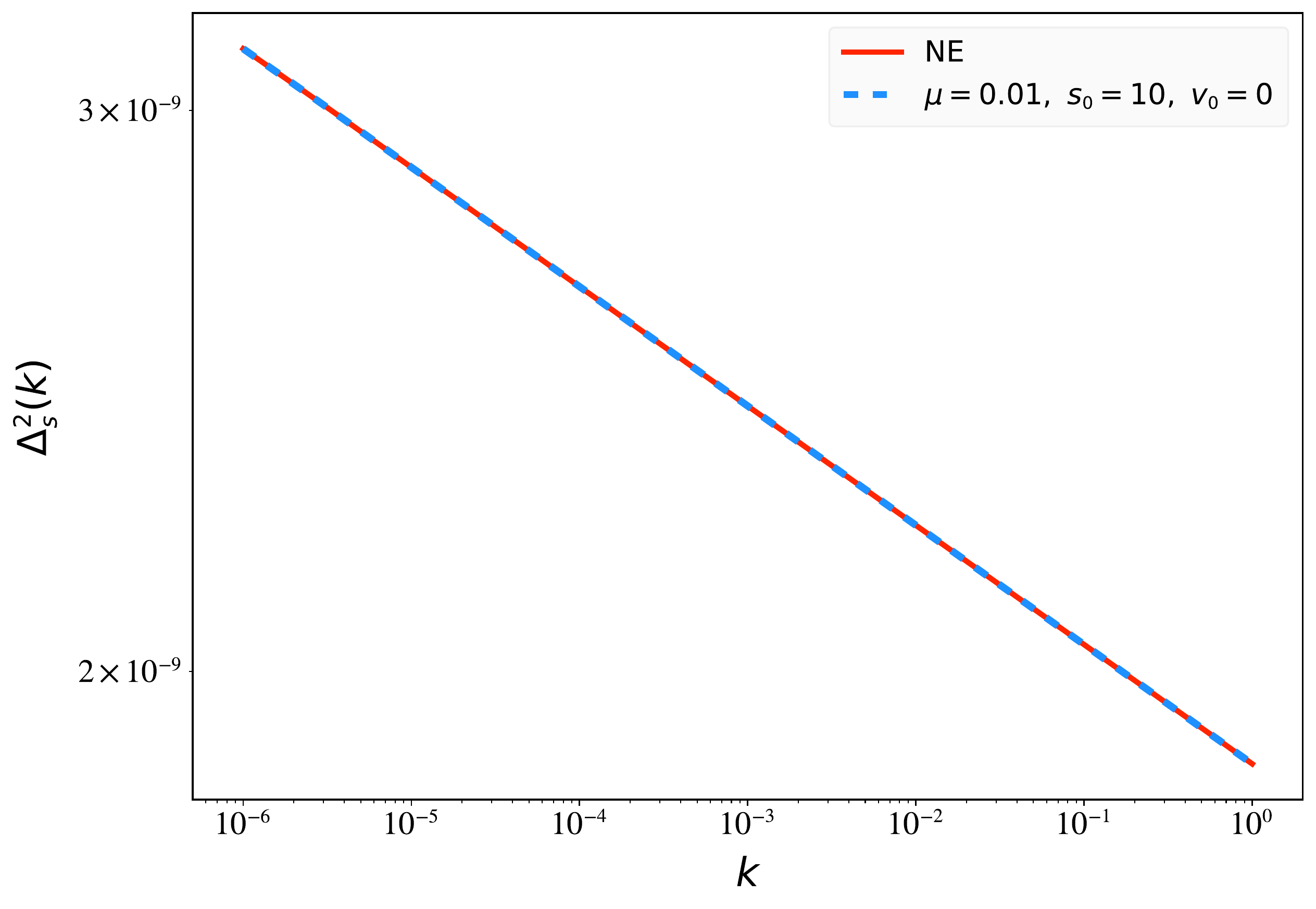}
    \vspace{0.1cm}
  \end{subfigure} 
  \begin{subfigure}[b]{0.62\linewidth}
  \centering
    \includegraphics[width=\linewidth]{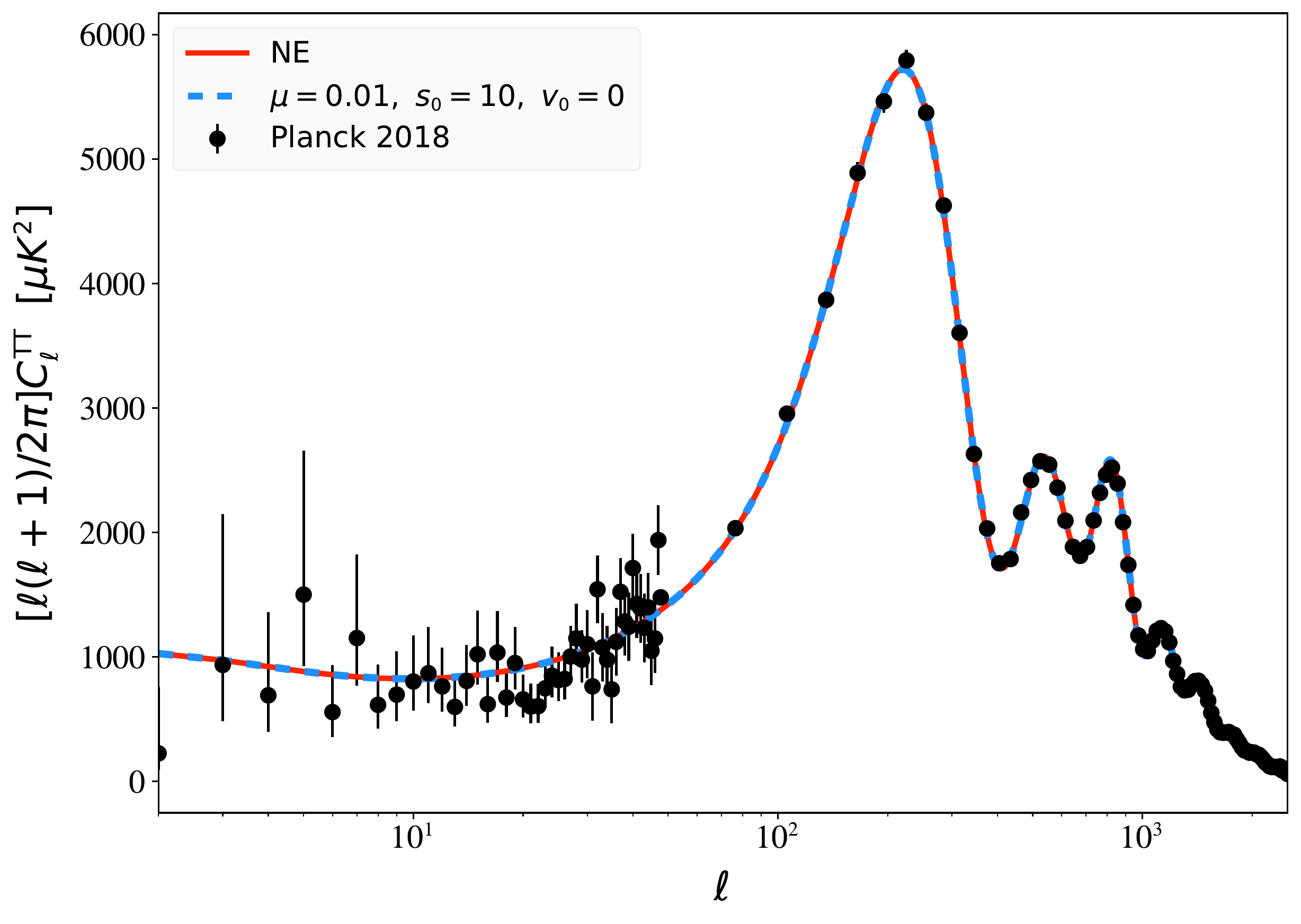}
    \vspace{0.1cm}
  \end{subfigure}
  \begin{subfigure}[b]{0.62\linewidth}
    \centering
    \includegraphics[width=\linewidth]{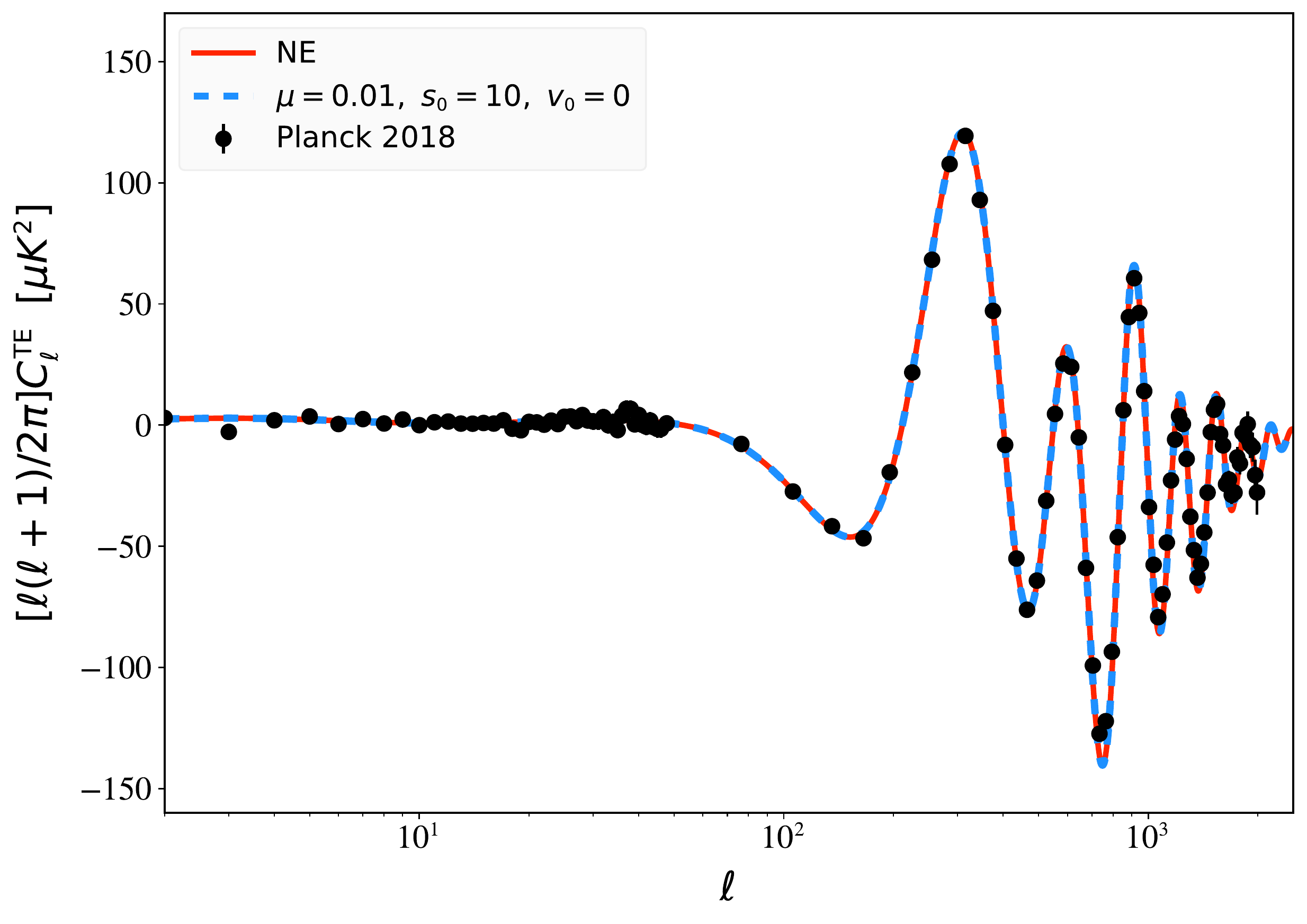}
  \end{subfigure}
  \caption{The primordial power spectrum (top), the unlensed TT power spectrum (middle), and the unlensed TE power spectrum (bottom) for $\mu=0.01$, $s_0=10$, and $v_0 =0$. In all cases, the power spectra are compared with the non-entangled versions originating from the use of a Bunch-Davies state in the models. The angular power spectra (middle and bottom) are also compared with the CMB data from Planck.} 
\label{{fig:ClandPkmu0.01nov0}}
\end{figure}

\begin{figure}[hbtp]
\centering
  \begin{subfigure}[b]{0.62\linewidth}
    \includegraphics[width=\linewidth]{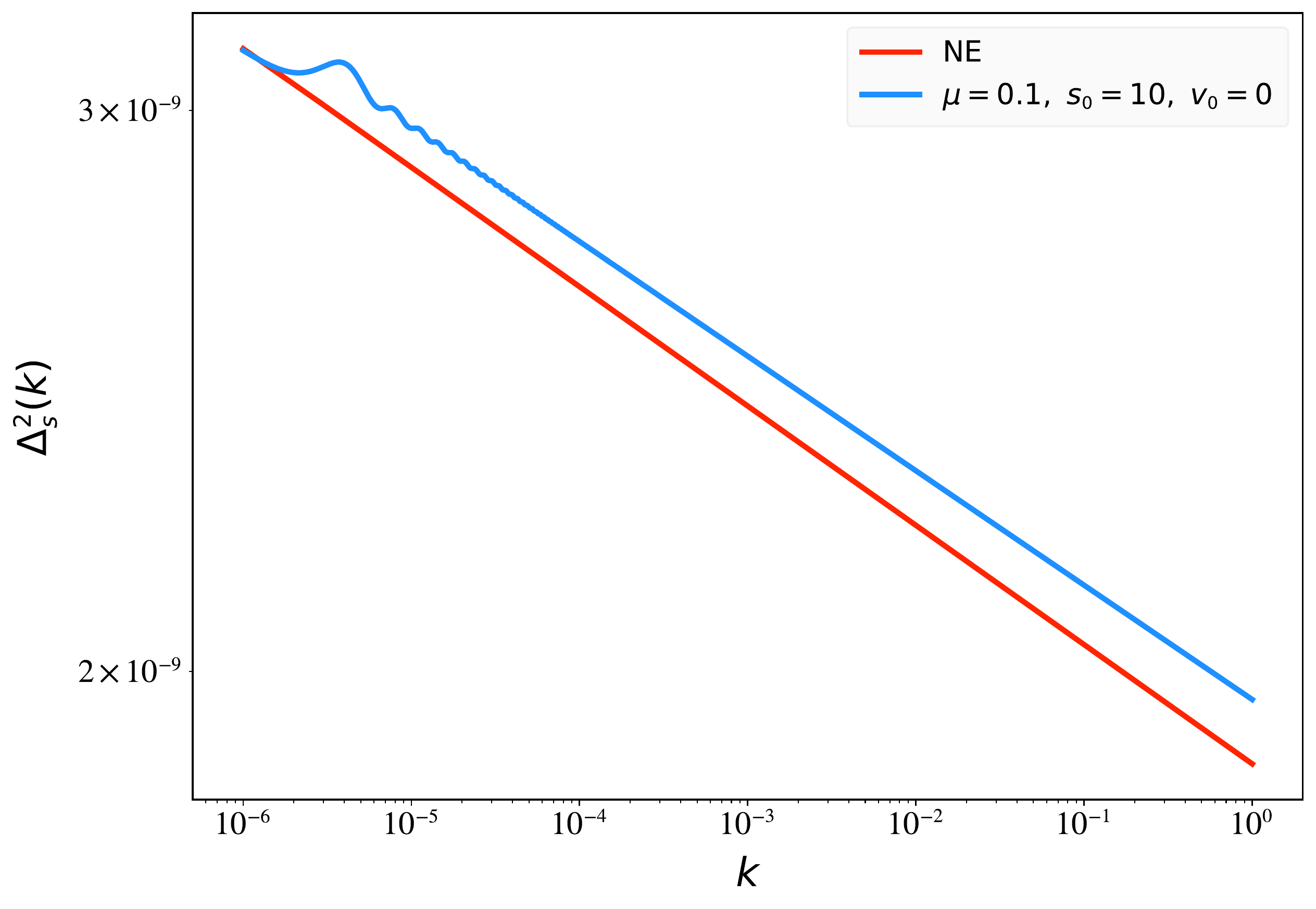}
    \vspace{0.1cm}
  \end{subfigure}
  \begin{subfigure}[b]{0.62\linewidth}
    \includegraphics[width=\linewidth]{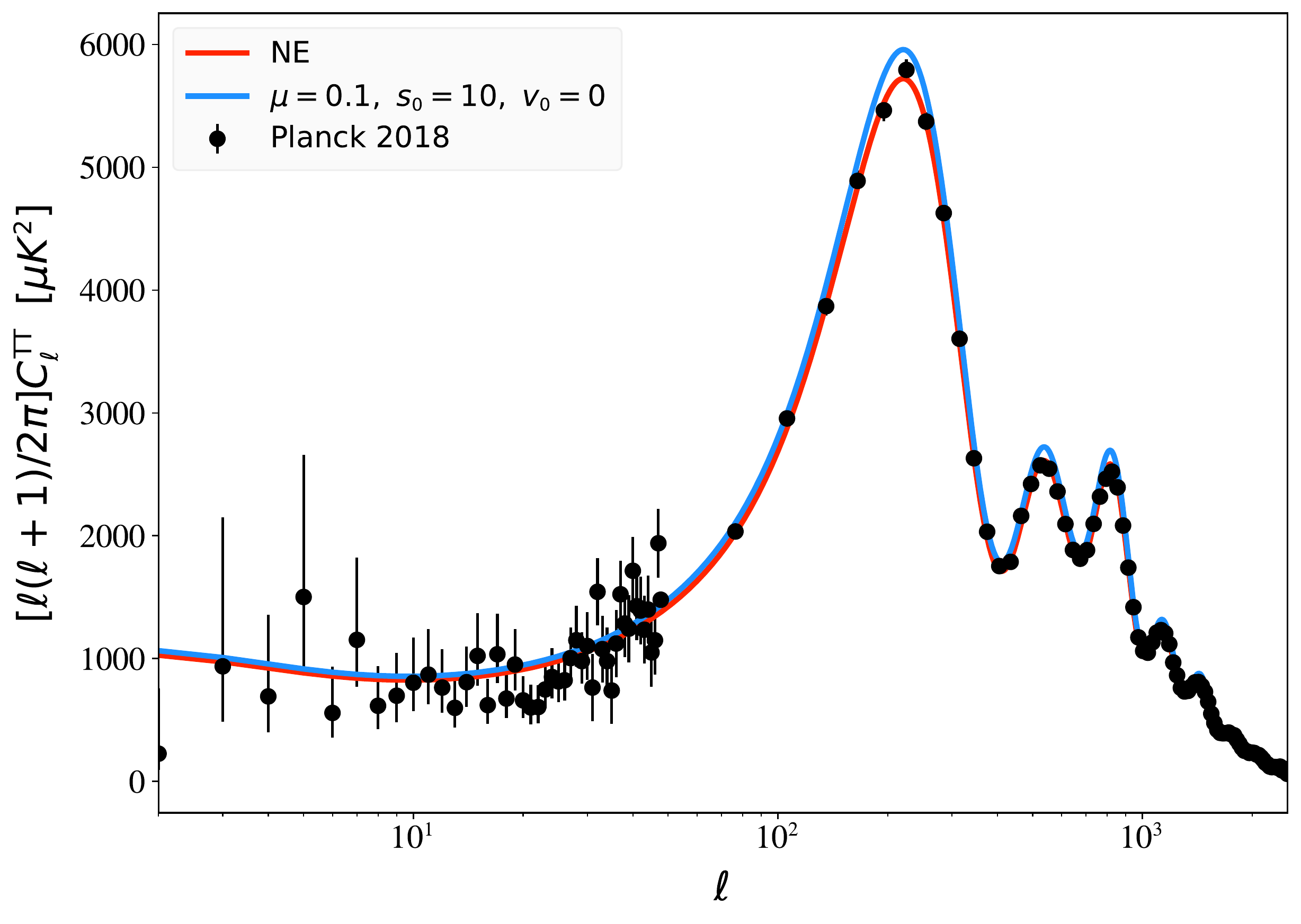}
    \vspace{0.1cm}
  \end{subfigure}
  \vspace{0.1cm}  
  \begin{subfigure}[b]{0.62\linewidth}
    \includegraphics[width=\linewidth]{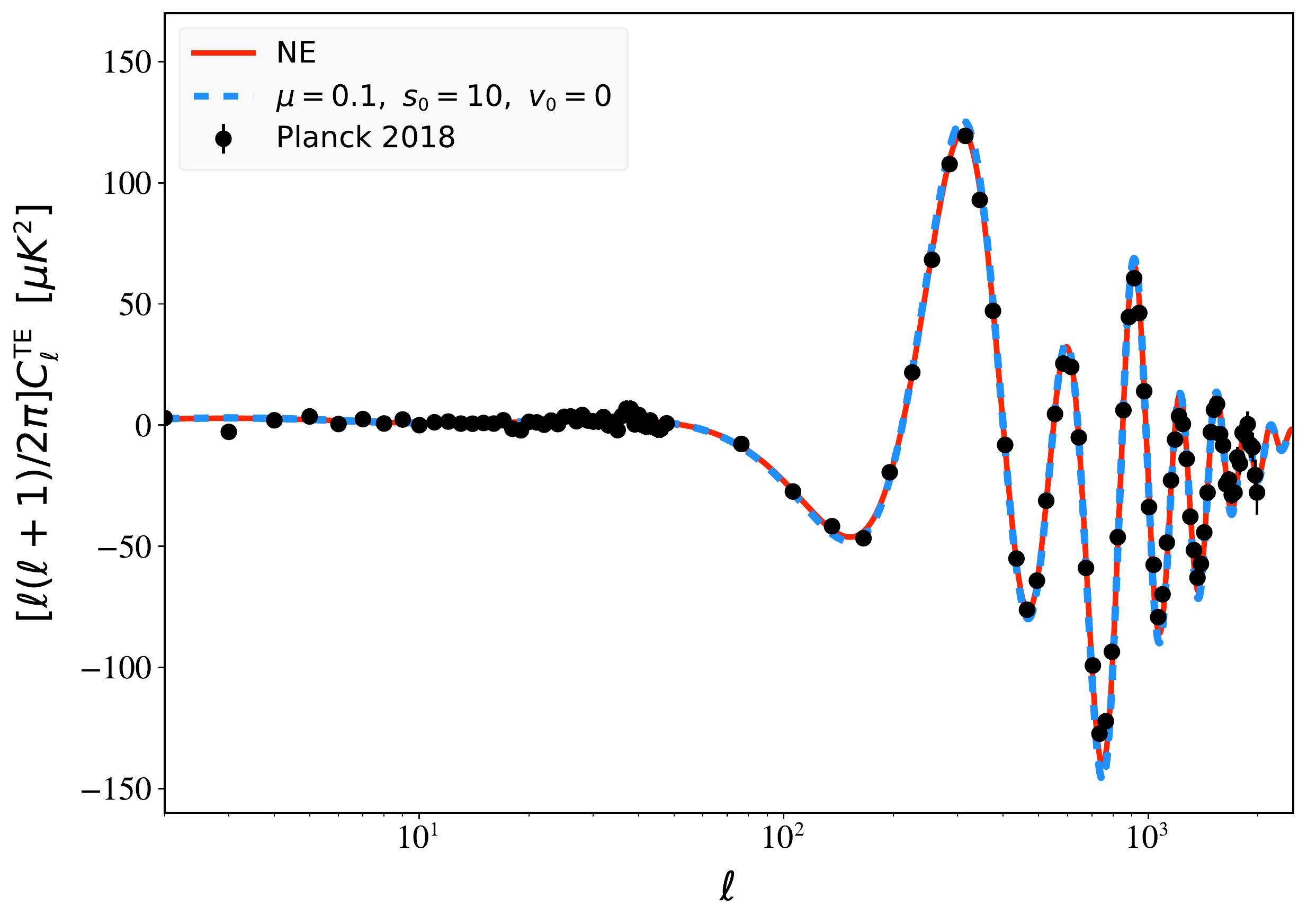}
  \end{subfigure}
  \caption{The primordial power spectrum (top), the unlensed TT power spectrum (middle), and the unlensed TE power spectrum (bottom) for $\mu=0.1$, $s_0=10$, and $v_0 =0$. As in Figure \ref{{fig:ClandPkmu0.01nov0}}, non-entangled power spectra are also displayed (all subfigures) in addition to the CMB data from Planck (middle and bottom figures only).}
\label{{fig:ClandPkmu0.1nov0}}
\end{figure}

\begin{figure}[hbtp]
\centering
  \begin{subfigure}[b]{0.62\textwidth}
    \includegraphics[width=\textwidth]{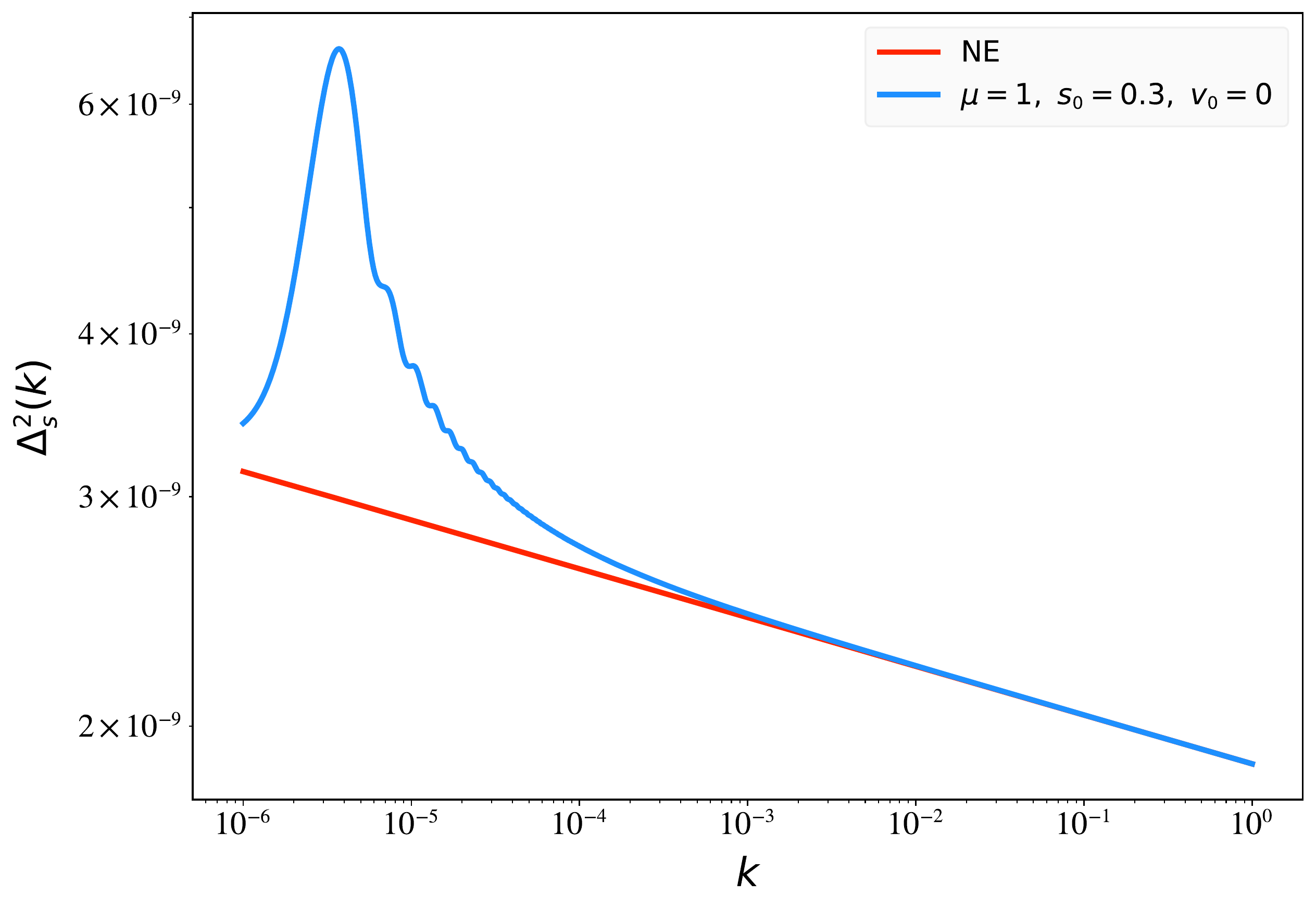}
     \vspace{0.1cm} 
  \end{subfigure}
  \begin{subfigure}[b]{0.62\textwidth}
    \includegraphics[width=\textwidth]{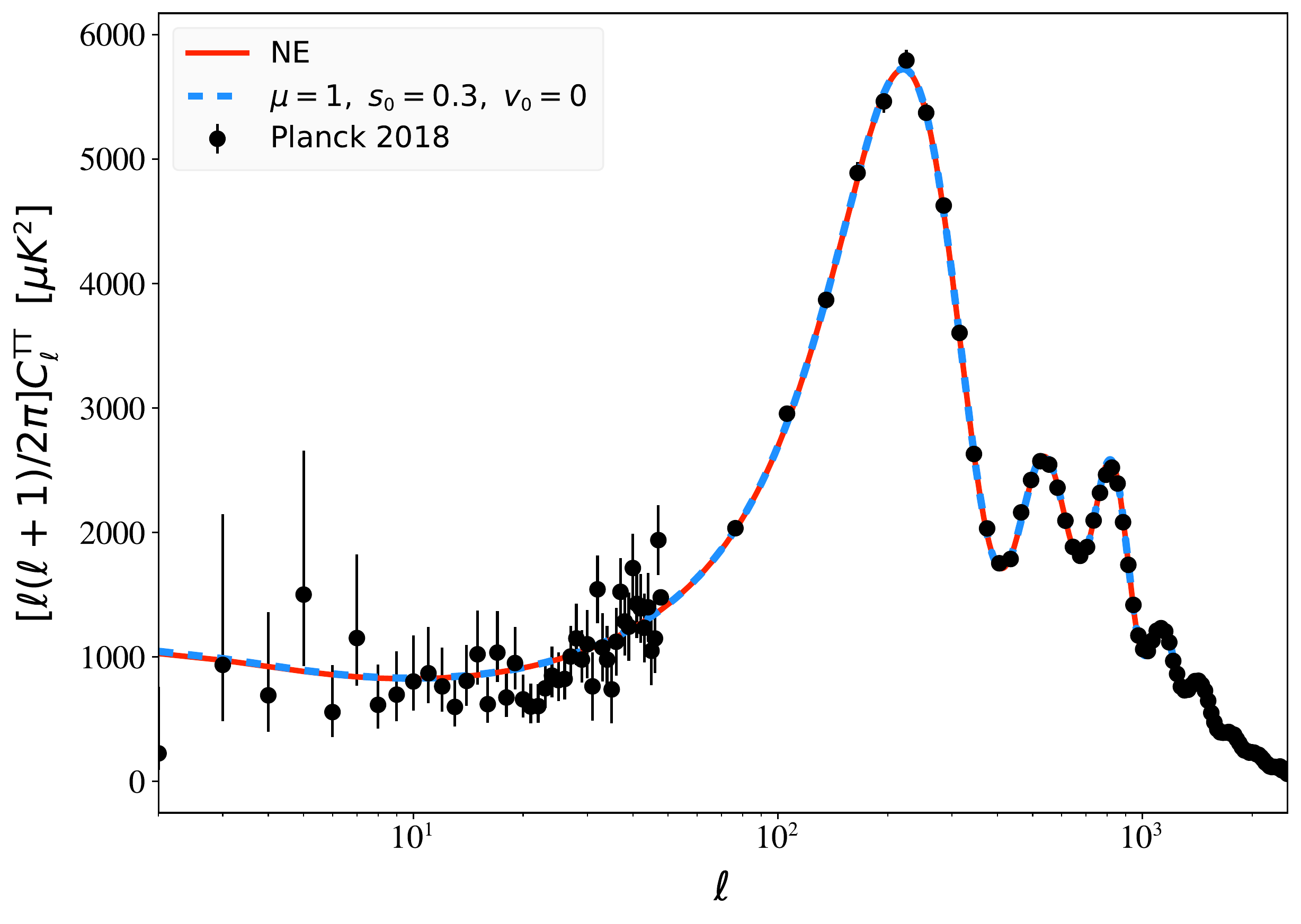}
    \vspace{0.1cm}
  \end{subfigure}
    \begin{subfigure}[b]{0.62\textwidth}
    \includegraphics[width=\textwidth]{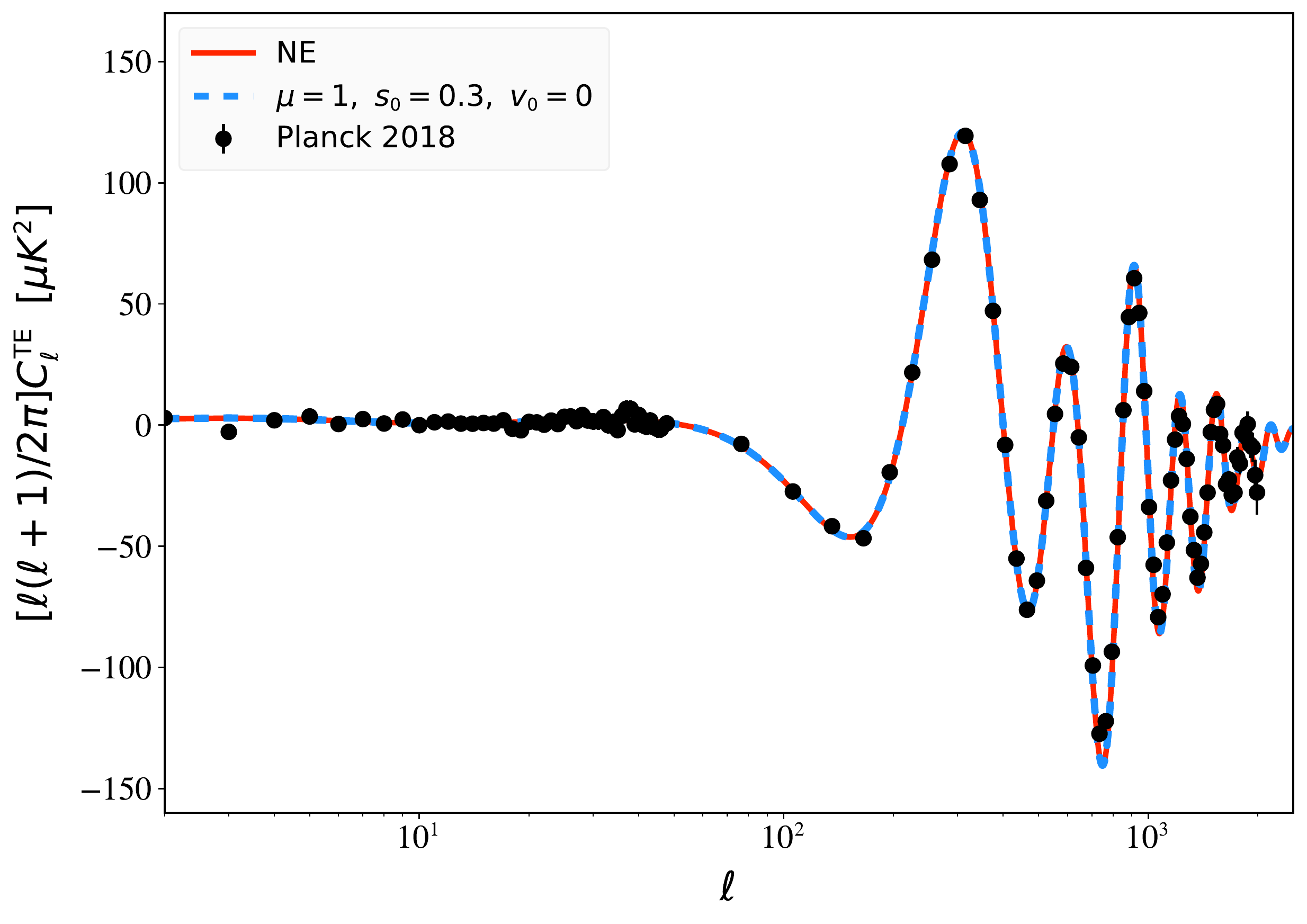}
  \end{subfigure}
\caption{The primordial power spectrum (top), the unlensed TT power spectrum (middle), and the unlensed TE power spectrum (bottom) for $\mu=1$, $s_0=0.3$, and $v_0 =0$. All the subfigures presented are similar to those in Figures \ref{{fig:ClandPkmu0.01nov0}} and \ref{{fig:ClandPkmu0.1nov0}} but for a new set of parameter values $\mu$ and $s_0$.}
\label{{fig:ClandPkmu1nov0}}
\end{figure}

\begin{figure}[hbtp]
\centering
  \begin{subfigure}[b]{0.62\linewidth}
    \includegraphics[width=\linewidth]{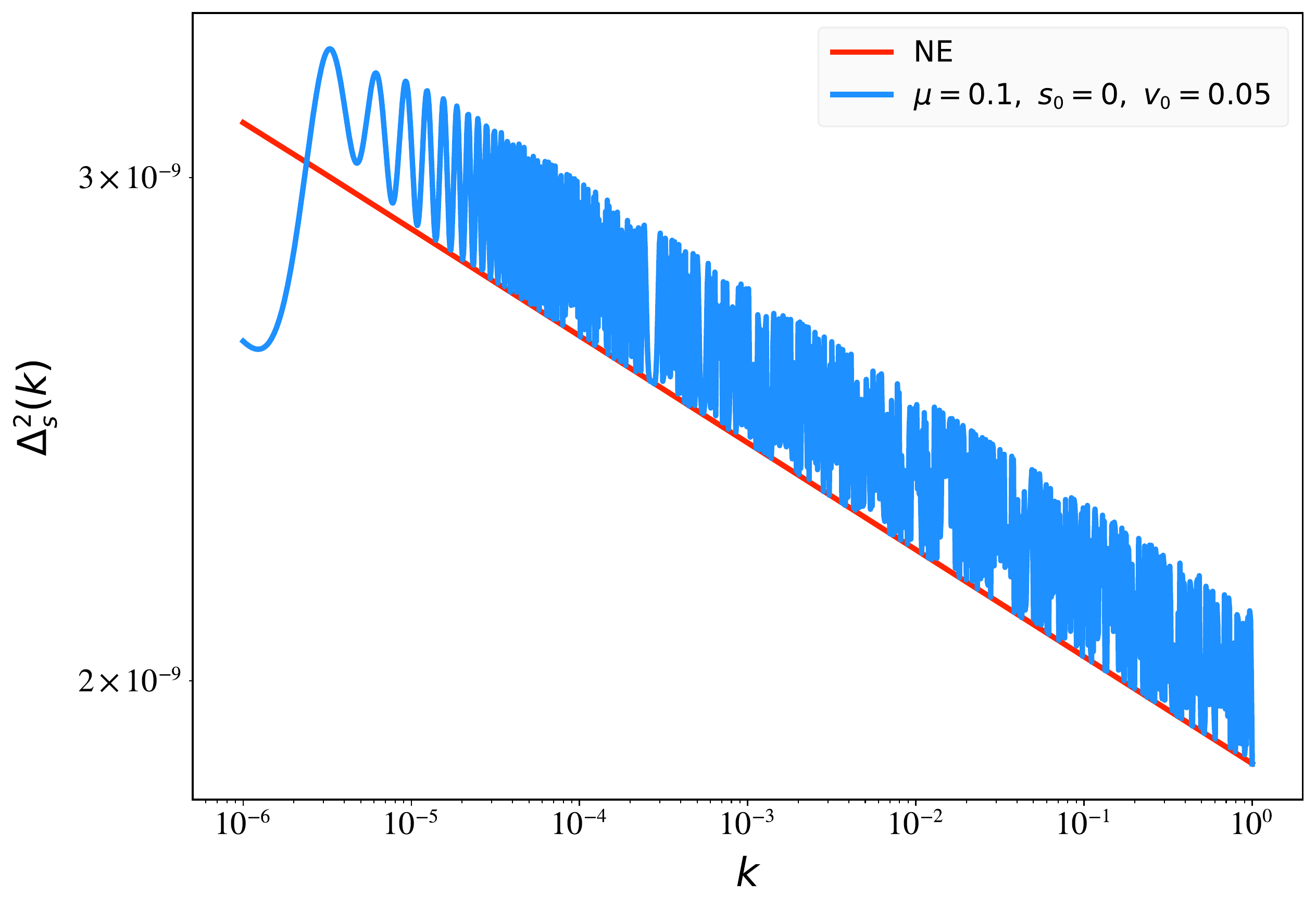}
    \vspace{0.1cm}
  \end{subfigure}
  \begin{subfigure}[b]{0.62\linewidth}
    \includegraphics[width=\linewidth]{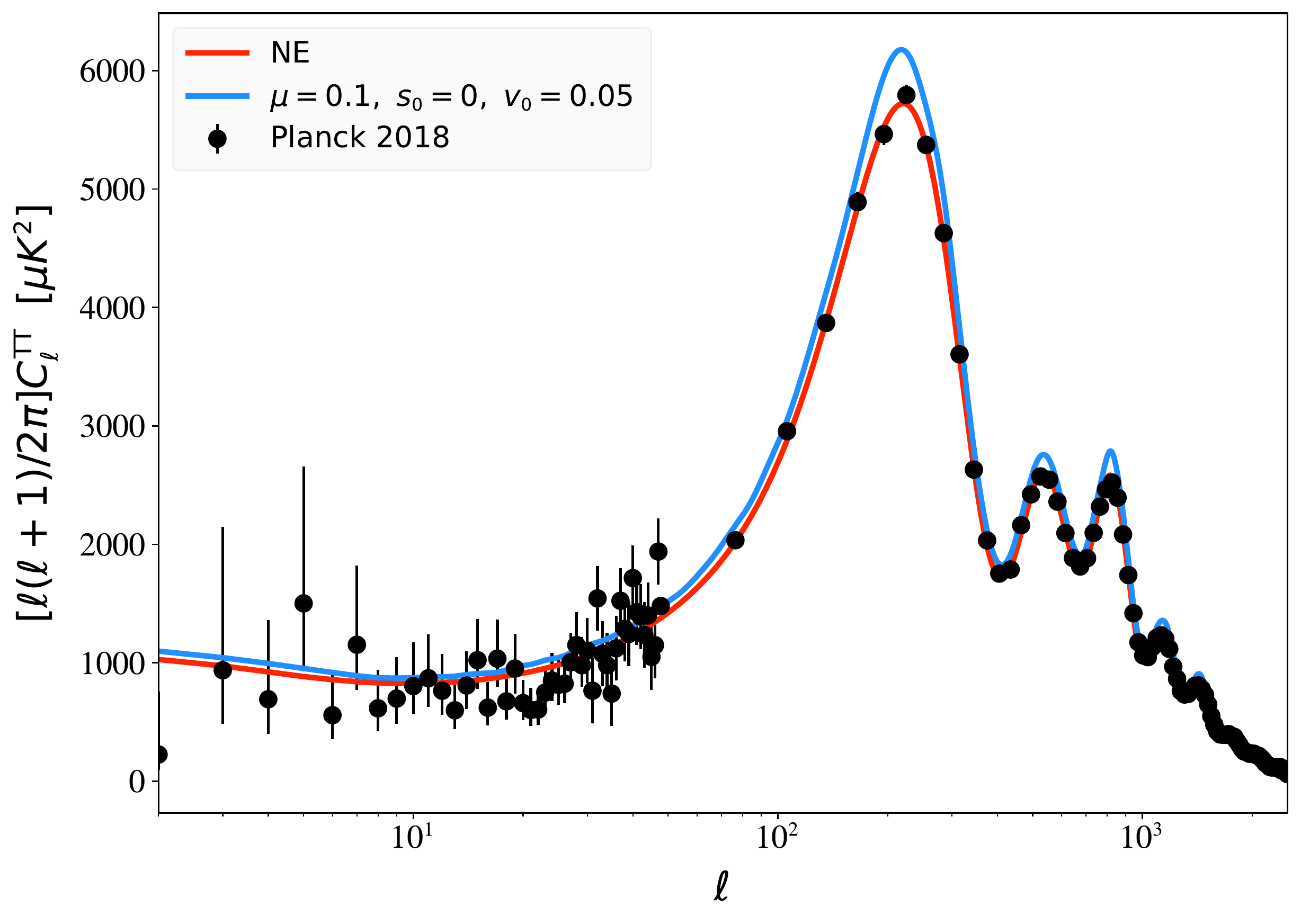}
    \vspace{0.1cm}
  \end{subfigure}
  \begin{subfigure}[b]{0.62\linewidth}
    \includegraphics[width=\linewidth]{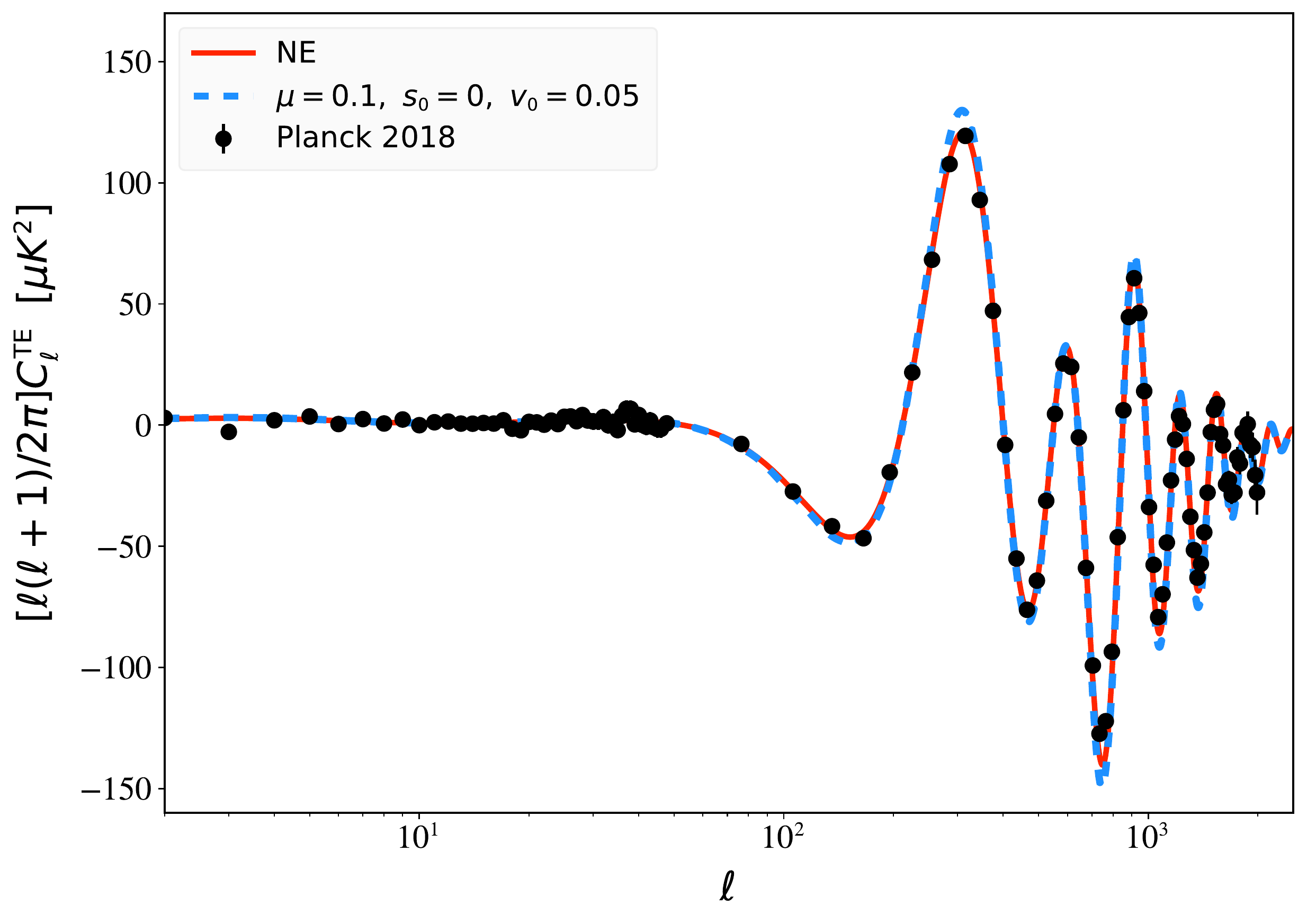}
  \end{subfigure}
\caption{The primordial power spectrum (top), the unlensed TT power spectrum (middle), and the unlensed TE power spectrum (bottom) for $\mu=0.1$, $s_0=0$, and $v_0 =0.05$. All the curves shown are comparable to the ones displayed in Figures \ref{{fig:ClandPkmu0.01nov0}}, \ref{{fig:ClandPkmu0.1nov0}}, and \ref{{fig:ClandPkmu1nov0}} for different choices of parameter values $\mu$, $s_0$, and $v_0$.}
\label{{fig:ClandPkmu0.1nos0}}
\end{figure}

Figure \ref{{fig:ClandPkmu0.1nov0}} provides the same information as the previous figure, except that the entangled case has the following parameters: $\mu=0.1$, $s_0=10$, and $v_0 =0$. This is the higher-mass version of Figure \ref{{fig:ClandPkmu0.01nov0}} and the differences between the entangled and non-entangled cases are more apparent here. The oscillations in the primordial spectrum for the entangled case quickly damp out and no real oscillations are apparent in the resulting TT spectra. There is an enhancement of power in the peaks of the TT spectrum for the entangled case that puts this set of parameters just outside observational bounds by eye (given the constraints of our analysis). The rest of the curve, however, is well within the error bars of the Planck data. The TE power spectra are barely distinguishable from one another. A full parameter analysis, which we postpone to future work, would be able to determine whether this set of parameter values is truly ruled out by the data or not.

The results in Figure \ref{{fig:ClandPkmu1nov0}}, which are the same comparisons as the previous two figures for the entangled case $\mu=1$, $s_0=0.3$, and $v_0 =0$, are curious. Despite the large enhancement of power for low $k$ in the primordial power spectrum in the entangled case, the resulting TT and TE spectra are indistinguishable by eye from the standard non-entangled result. Unlike the outcomes in Figure \ref{{fig:ClandPkmu0.01nov0}}, these results are a more dramatic instance of asking whether Planck data can distinguish the BD state from an entangled state. The primordial spectrum here is noticeably different in the entangled case, yet it seems to have no effect on the TT and TE spectra. It was the results of this set of parameters that provoked us to explore changing the onset of entanglement---to see what would happen if we shift features around to effectively higher $k$ values---which we explore in section \ref{subsec:k0shift}.

Lastly, Figure \ref{{fig:ClandPkmu0.1nos0}} explores an entangled case with a non-zero initial velocity. For the parameters $\mu=0.1$, $s_0=0$, and $v_0 =0.05$, the effect is to have high frequency oscillations for the majority of the observable $k$ range in the entangled primordial power spectrum, which translates to a TT spectrum that sits above the non-entangled case. While the distance from the non-entangled TT spectrum is not constant---so one might be able to argue the presence of some oscillations---overall the oscillations from the entangled primordial spectrum appear averaged over. More of the entangled TT spectrum is outside the bounds of Planck compared to Figure \ref{{fig:ClandPkmu0.1nov0}}. Furthermore, the TE spectrum overshoots several peaks. This is likely a set of parameters that a full parameter estimation would be able to reject.

\subsection{Axion Spectator Field}
\label{subsec:axion}
Axions in the early universe are well motivated (see e.g.~\cite{Marsh:2015xka}), so a spectator field with an axion-like potential also merits consideration. We take the potential to be of the form 
\begin{equation}
\label{eq;axionpot}
V(\sigma) = \Lambda^4 \left(1-\cos\left(\frac{\sigma}{f_a}\right)\right),
\end{equation}
where $\Lambda^4$ is the energy density associated with the potential and $f_a$ the axion decay constant. Note that this need not be the QCD axion. It follows that the dimensionless potential $\bar{V}(s)$ is then given by:
\begin{equation}
\label{eq:dimensionlessaxionpot}
\bar{V}(s) = 1 - \cos{ ( s \slash \tilde{f}_{a} )},
\end{equation}
where $\tilde{f}_a=f_a\slash M_{\rm Pl}$ and the dimensionless mass squared term corresponds to $\mu^2= \Lambda^4\slash (M_{\rm Pl}^2 H_{\rm dS}^2)$.

In addition to varying $\mu$, $s_{0}$, and $v_{0}$, the behavior of the spectator can also change depending on the choice of $\tilde{f}_{a}$. We illustrate some of these variations in the subsequent figures. Since a full analysis of all possible initial conditions is beyond the scope of this paper, we restrict ourselves to parameters that provide interesting behaviors distinct from the free massive scalar case. 
\begin{figure}[hbtp]
    \centering
    \begin{subfigure}[b]{0.62\linewidth}
        \centering
        \includegraphics[width=\linewidth]{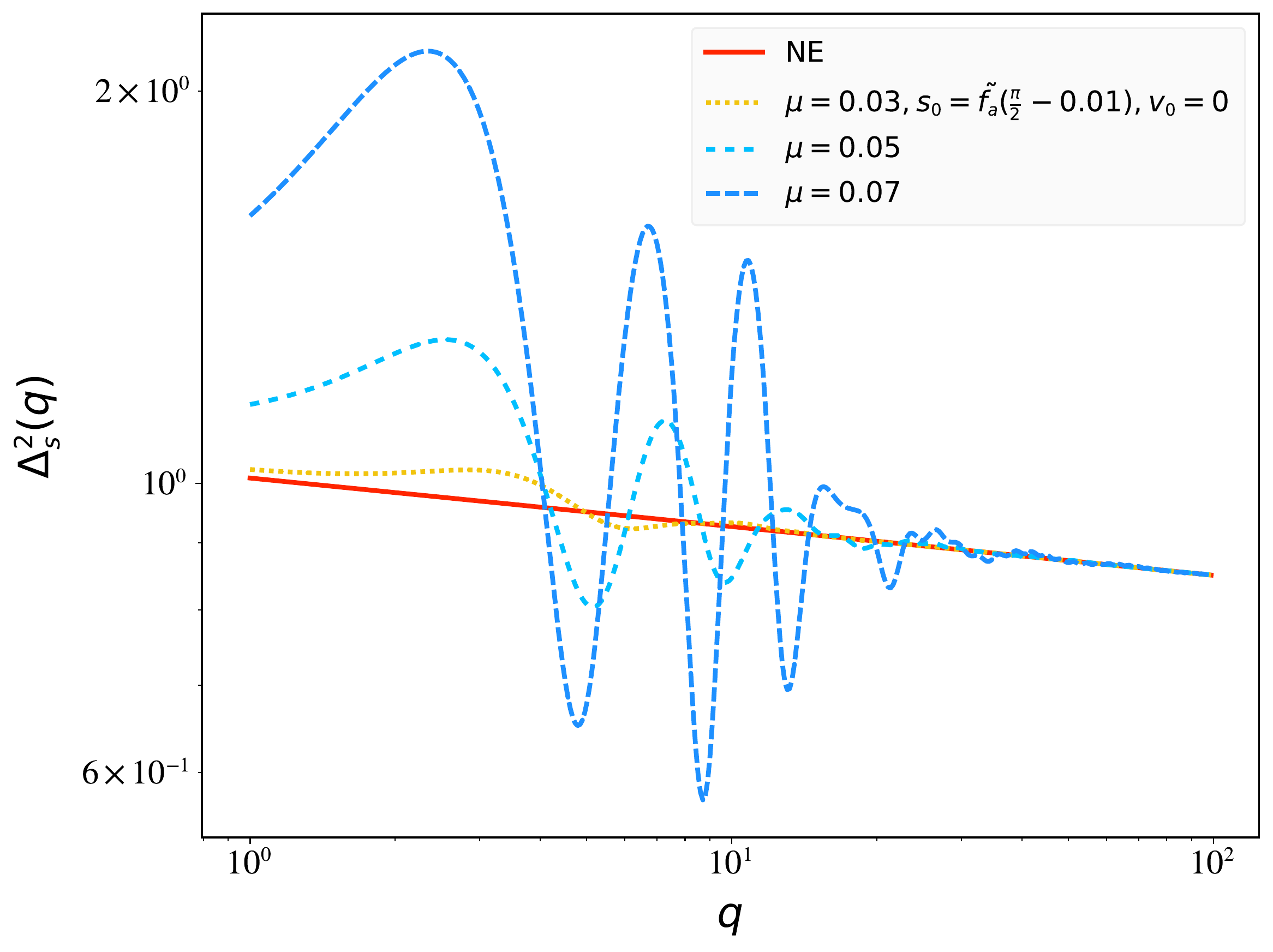}
        \subcaption{$\tilde{f}_{a}=0.01$}
        \vspace{0.1cm}
    \end{subfigure}    
    \begin{subfigure}[b]{0.62\linewidth}
        \centering
        \includegraphics[width=\linewidth]{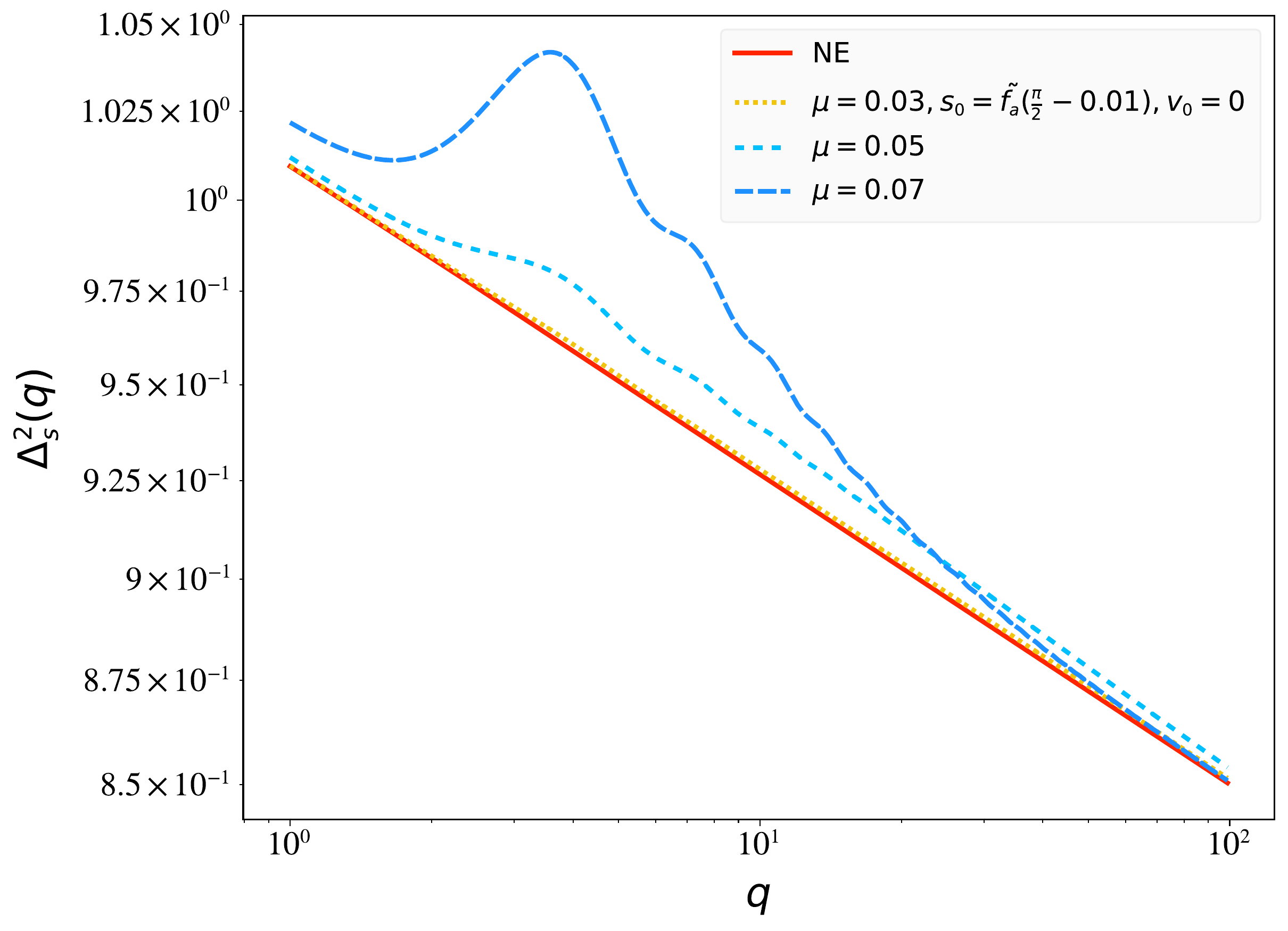}
        \subcaption{$\tilde{f}_{a} = 0.05$}
    \end{subfigure}
    \caption{Log-log plots of the power spectrum $\Delta_s^2$ plotted in units of $A_s$ as a function of $q=k\slash {\mathcal H}_0$. Here, we compare the impact of adjusting $\tilde{f}_{a}$ for the axion-like potential on a variety of masses, given the initial condition $s_0= \tilde{f_{a}} (\frac{\pi}{2} - 0.01) $ and $v_0 =0$. We take $\tilde{f}_{a} = 0.01$ in (a) and $\tilde{f}_{a} = 0.05$ in (b).}
  \label{fig:axionfacompare}
\end{figure}

In Figure \ref{fig:axionfacompare}, we see different patterns of oscillations for the same initial condition of $s_{0} = \tilde{f_{a}} (\frac{\pi}{2} - 0.01)$ and $v_{0} =0$, depending on the choice of $\mu$ and $\tilde{f_{a}}$. For the choice $\tilde{f_{a}} = 0.01$ (Figure \ref{fig:axionfacompare}a), we see higher values of $\mu$ generate more oscillations with a larger amplitude for low $q$. However, in all cases shown these oscillations damp out and coalesce so that the high $q$ (or $k$) primordial power spectrum is indistinguishable from the non-entangled case (see also Figure \ref{{fig:ClandPkaxion}}). In contrast, for the case $\tilde{f_{a}} = 0.05$ shown in Figure \ref{fig:axionfacompare}b (for the same masses and initial condition as Figure \ref{fig:axionfacompare}a) we observe a different behavior. After an initial enhancement of power for low $q$, oscillations damp out and decay in a manner reminiscent of Figure \ref{fig:zerovelocitycase}c. We see larger initial enhancements of power for larger masses.

Since the behaviors in Figure \ref{fig:axionfacompare}a are more distinct from the free massive scalar results, we also investigated the effect of adding a small non-zero $v_{0}$ to those initial conditions. The result is shown in Figure \ref{{fig:axionv0compare}}. The pattern of large initial oscillations that damp out is still present, although their form is modified compared to Figure \ref{fig:axionfacompare}a. However, we notice that the cases shown all exhibit the same behavior at higher $q$ values, after the initial oscillations damp out, regardless of the value of $\mu$. These high $q$ oscillations have an approximately constant amplitude, and their location relative to the non-entangled case is similar to Figure \ref{{fig:ClandPkmu0.1nos0}}.
\begin{figure}[hbtp]
\centering
  \includegraphics[scale=0.4]{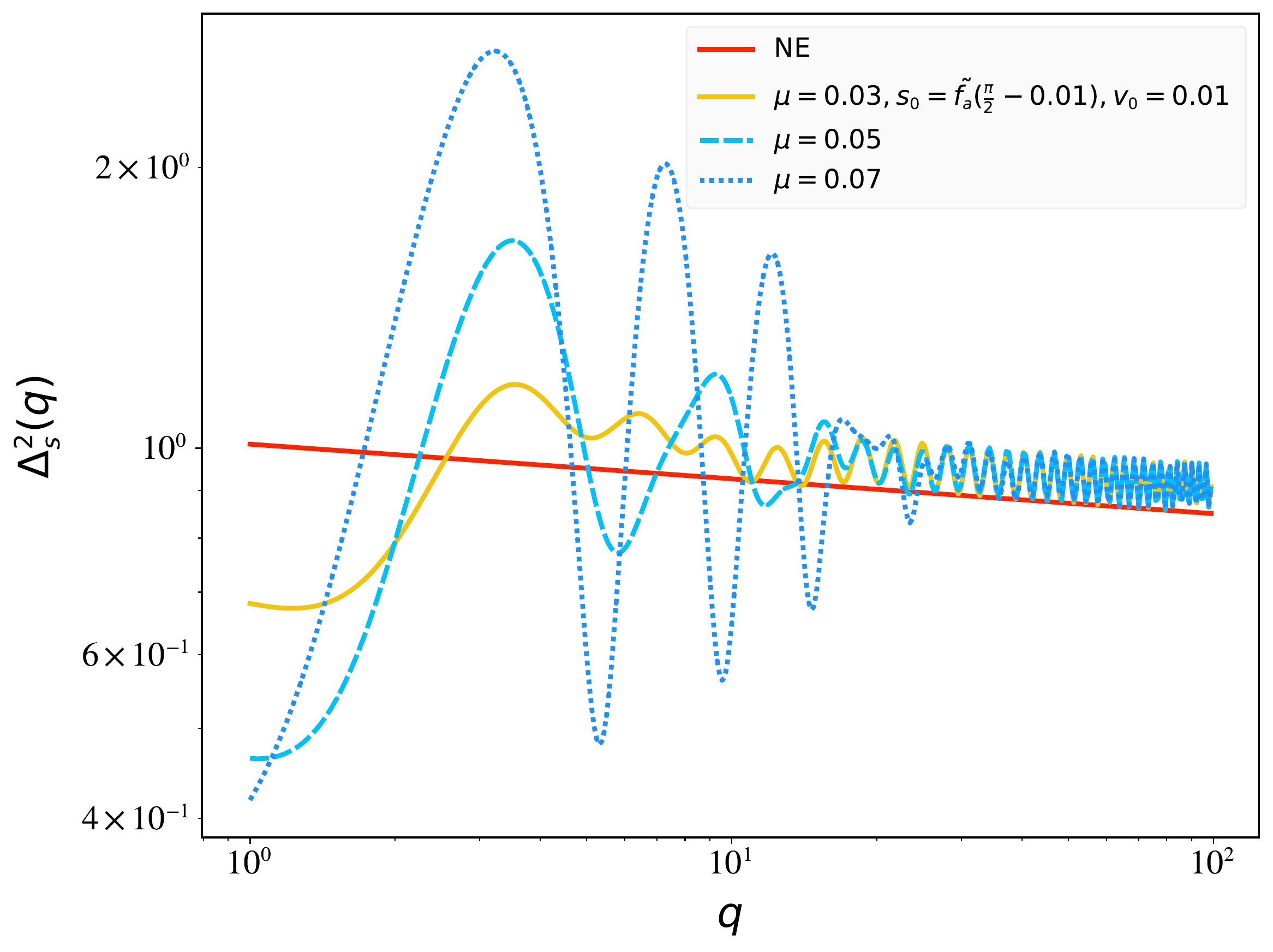}
\caption{Log-log plot of the power spectrum for $s_{0}=\tilde{f_{a}}(\frac{\pi}{2} - 0.01)$, $v_{0} = 0.01$, $\tilde{f_{a}} = 0.01$, and different choices of $\mu$.}
\label{{fig:axionv0compare}}
\end{figure}

\begin{figure}[hbtp]
\centering
  \begin{subfigure}[b]{0.6\linewidth}
  \centering
    \includegraphics[width=\linewidth]{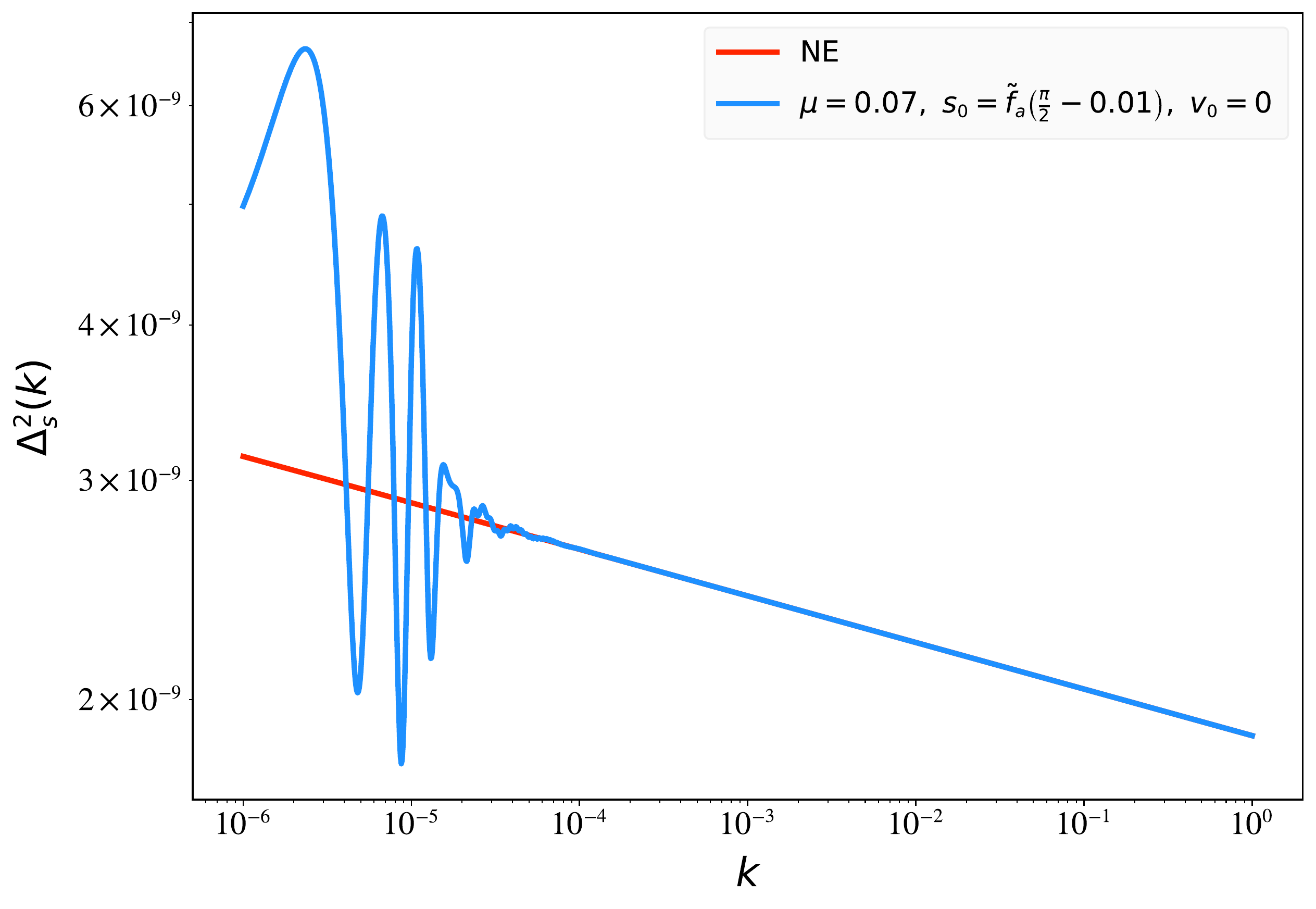}
  \vspace{0.1cm}
  \end{subfigure}  
  \begin{subfigure}[b]{0.6\linewidth}
  \centering
    \includegraphics[width=\linewidth]{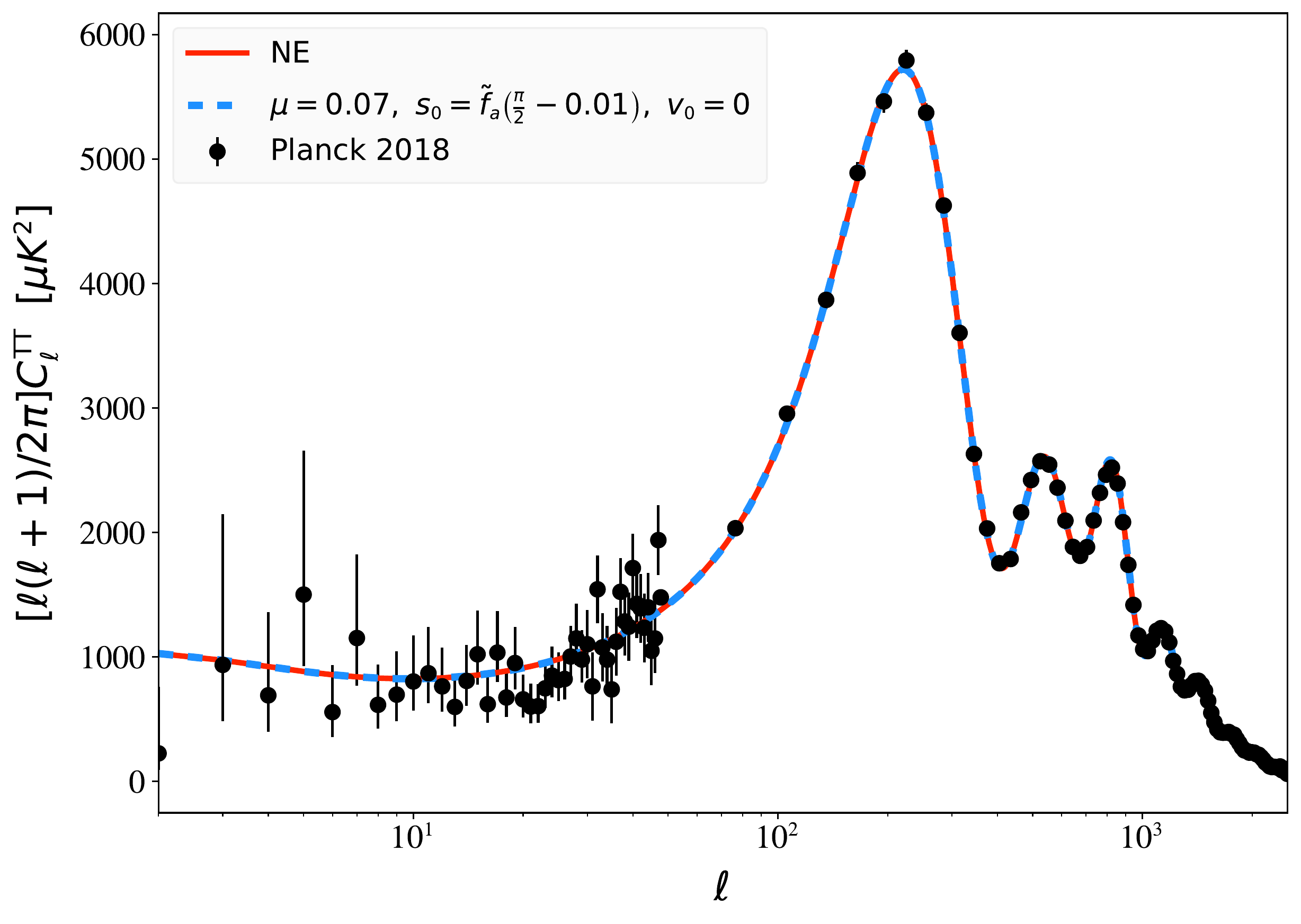}
  \vspace{0.1cm}
  \end{subfigure}
  \begin{subfigure}[b]{0.6\linewidth}
  \centering
    \includegraphics[width=\linewidth]{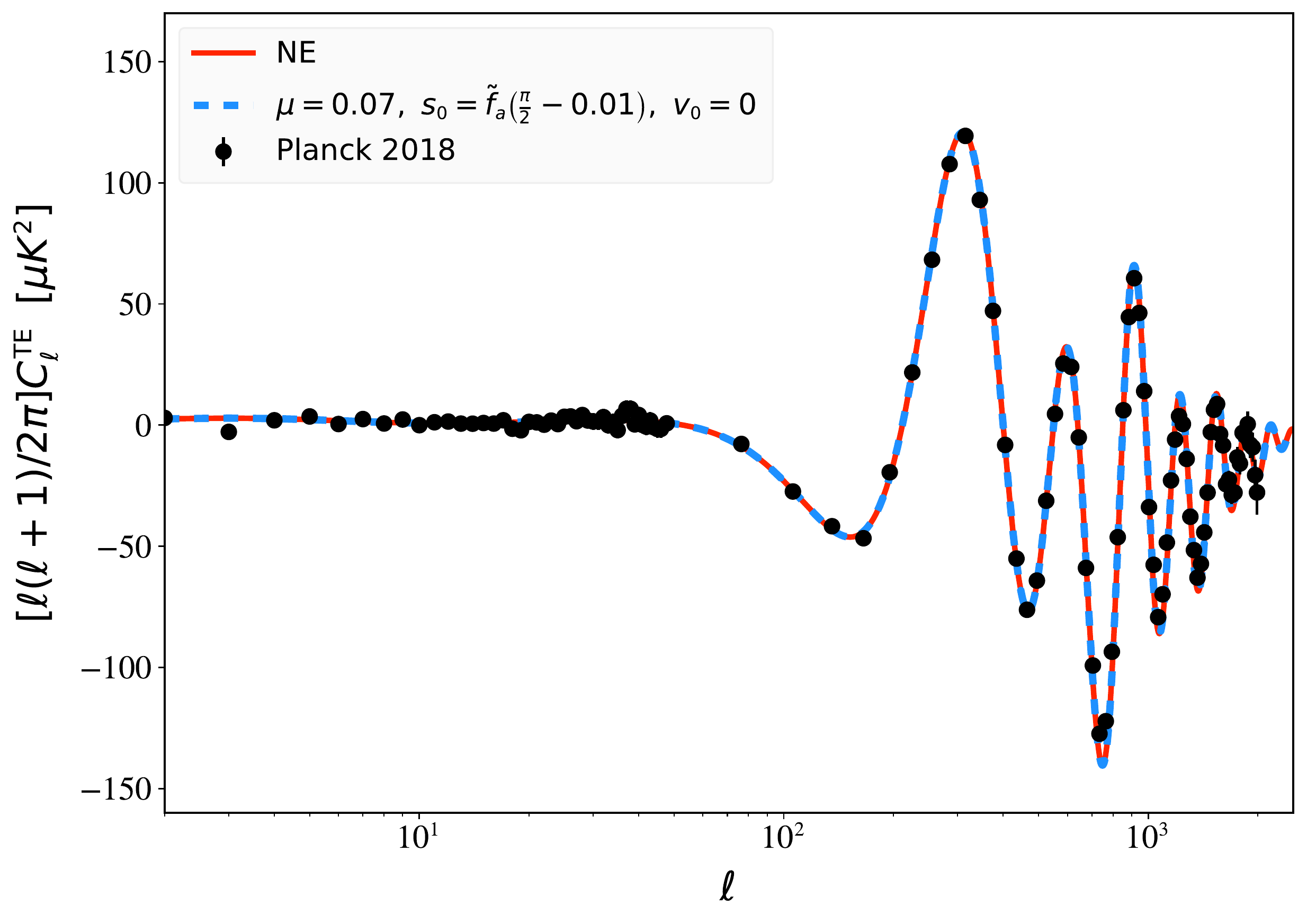}
  \end{subfigure}
\caption{The primordial power spectrum (top), the unlensed TT power spectrum (middle), and the unlensed TE power spectrum (bottom) for $\mu=0.07$, $s_0= \tilde{f_{a}} (\frac{\pi}{2} - 0.01) $, and $v_0 =0$, with $\tilde{f_{a}} = 0.01$. In all plots, the power spectra are compared with their Bunch-Davies counterparts. Additionally, the angular power spectra (middle and bottom) are compared with the CMB data from the Planck collaboration.}
\label{{fig:ClandPkaxion}}
\end{figure}

Figure \ref{{fig:ClandPkaxion}} investigates the effect of an axion-like potential with $\tilde{f_{a} }= 0.01$ for the entangled parameters $\mu=0.07$, $s_0= \tilde{f_{a}} (\frac{\pi}{2} - 0.01) $, and $v_0 =0$ on the CMB power spectra. Similar to what occurred with Figure \ref{{fig:ClandPkmu1nov0}}, there is no difference in the TT and TE spectra between the non-entangled case and entangled case, despite clear features in the primordial entangled power spectrum for these parameters. This is another case where BD and an entangled state both appear to be equally good fits to the Planck data, though obviously a full parameter estimation would be needed to push the case further. As with Figure \ref{{fig:ClandPkmu1nov0}}, the features in the primordial spectrum occur for low $k$, so in the next section we investigate what happens to the various power spectra if one changes the onset of entanglement.

\subsection{Shifting the Onset of Entanglement}
\label{subsec:k0shift}

As seen above, despite the presence of large changes in the primordial power spectrum for certain parameter choices, surprisingly only small differences were reflected in the CMB anisotropies. One conjecture is that this is due to the fact that we have set the onset of entanglement to coincide with the exiting of the largest length scale appearing on the CMB sky. To check this, we allow for the onset time, set by $\eta_0$, to correspond to smaller scales. Thus, we take $\eta_0$ to correspond to times well within the last 55 e-folds of inflation, with standard Bunch-Davies inflation being the initial condition. 

By the conversions in equation \eqref{eq:dimlesstimewave}, shifting the onset of entanglement (shifting $\eta_{0}$), translates to shifting $k_{0}$ in our code, where $k_{0}$ is the largest observable scale that will show evidence of entanglement. The results of our investigations for various values of $k_{0}$, effectively shifting non-standard features to higher $k$ values, are shown in Figures \ref{{fig:ClandPkmu0.1nos0k0shift}} - \ref{{fig:ClandPkaxionk0shift}}. 

\begin{figure}[hbtp]
\centering
  \begin{subfigure}[b]{0.62\linewidth}
    \centering
    \includegraphics[width=\linewidth]{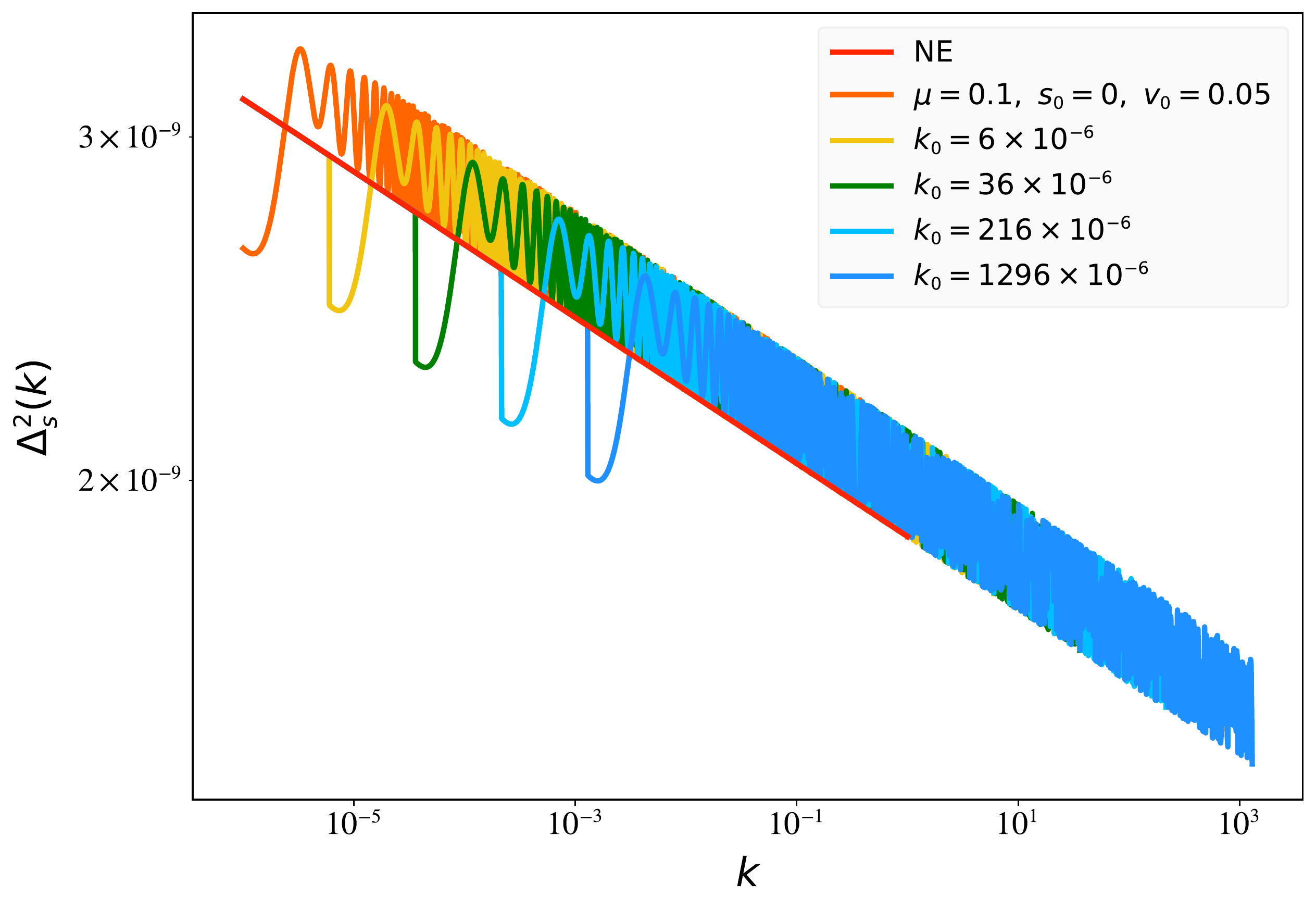}
    \vspace{0.1cm}
  \end{subfigure} 
  \begin{subfigure}[b]{0.62\linewidth}
    \centering
    \includegraphics[width=\linewidth]{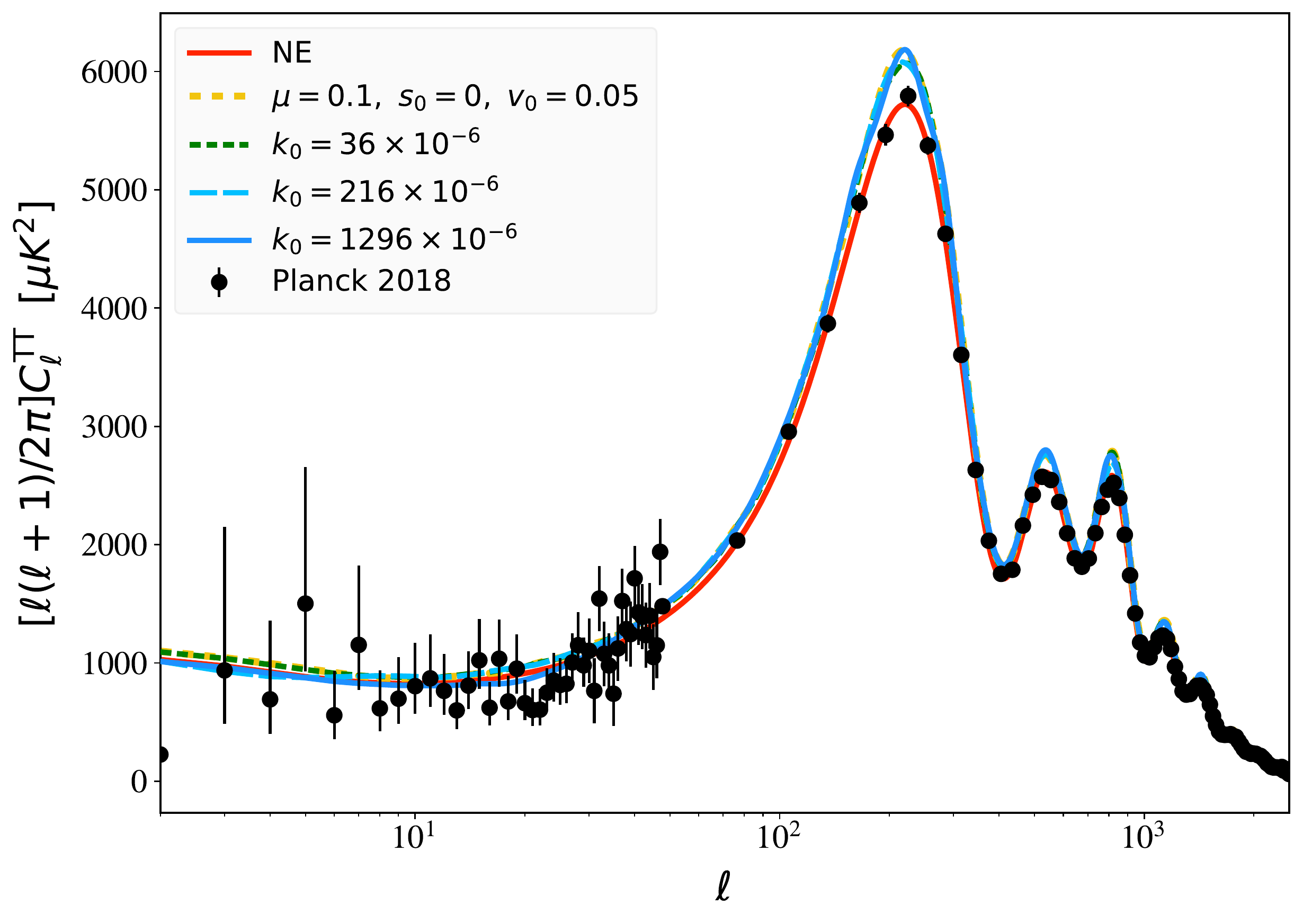}
    \vspace{0.1cm}
  \end{subfigure}
  \begin{subfigure}[b]{0.62\linewidth}
  \centering
    \includegraphics[width=\linewidth]{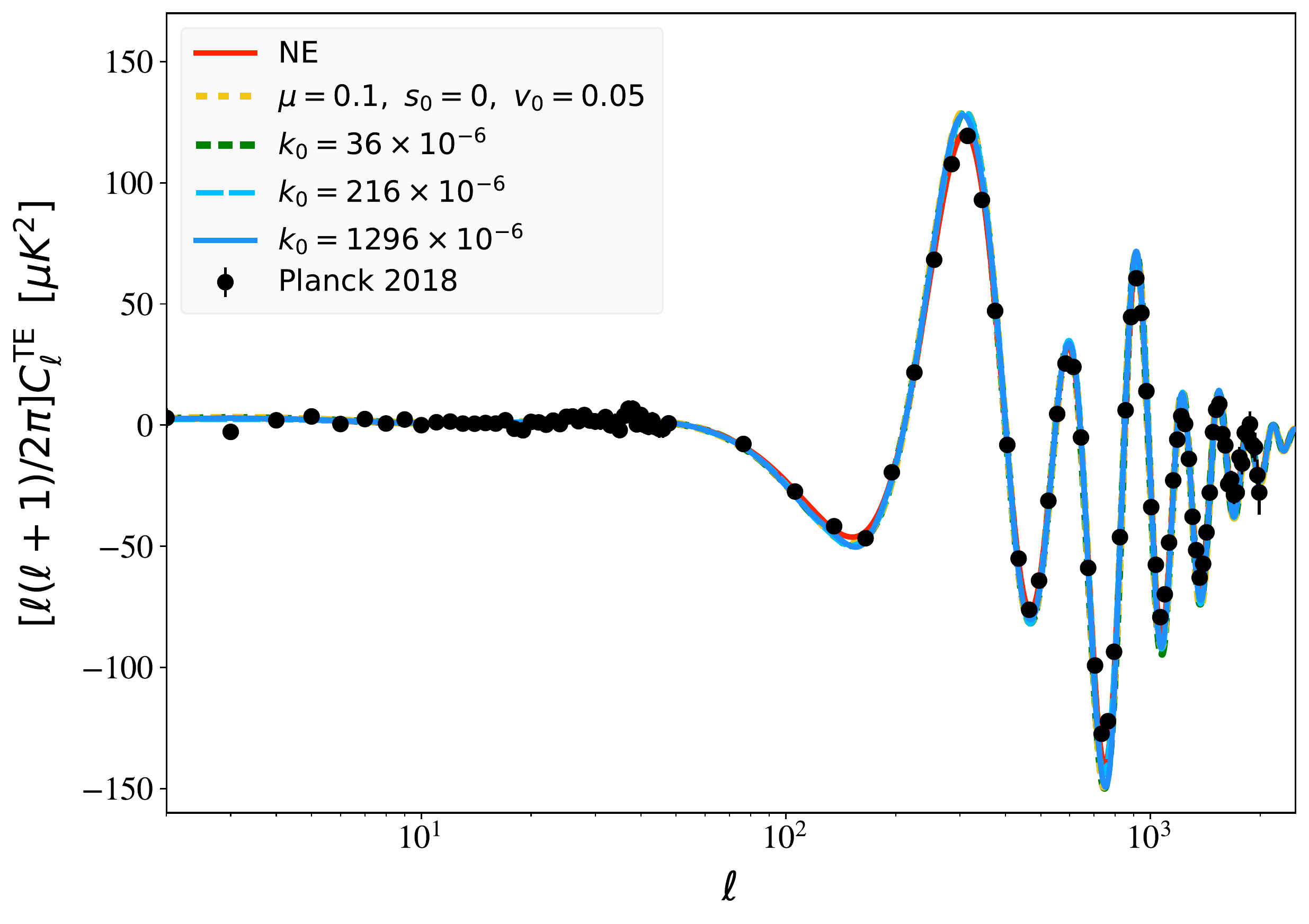}
  \end{subfigure}
\caption{The primordial power spectrum (top), the unlensed TT power spectrum (middle), and the unlensed TE power spectrum (bottom) for an entangled state involving a free massive scalar field with $\mu=0.1$, $s_0= 0$, and $v_0 =0.05$, for various values of $k_{0}$, compared with the non-entangled case (all subfigures) and CMB data from Planck (middle and bottom subfigures only).}
\label{{fig:ClandPkmu0.1nos0k0shift}}
\end{figure}

\begin{figure}[hbtp]
\centering
  \begin{subfigure}[b]{0.62\linewidth}
    \includegraphics[width=\linewidth]{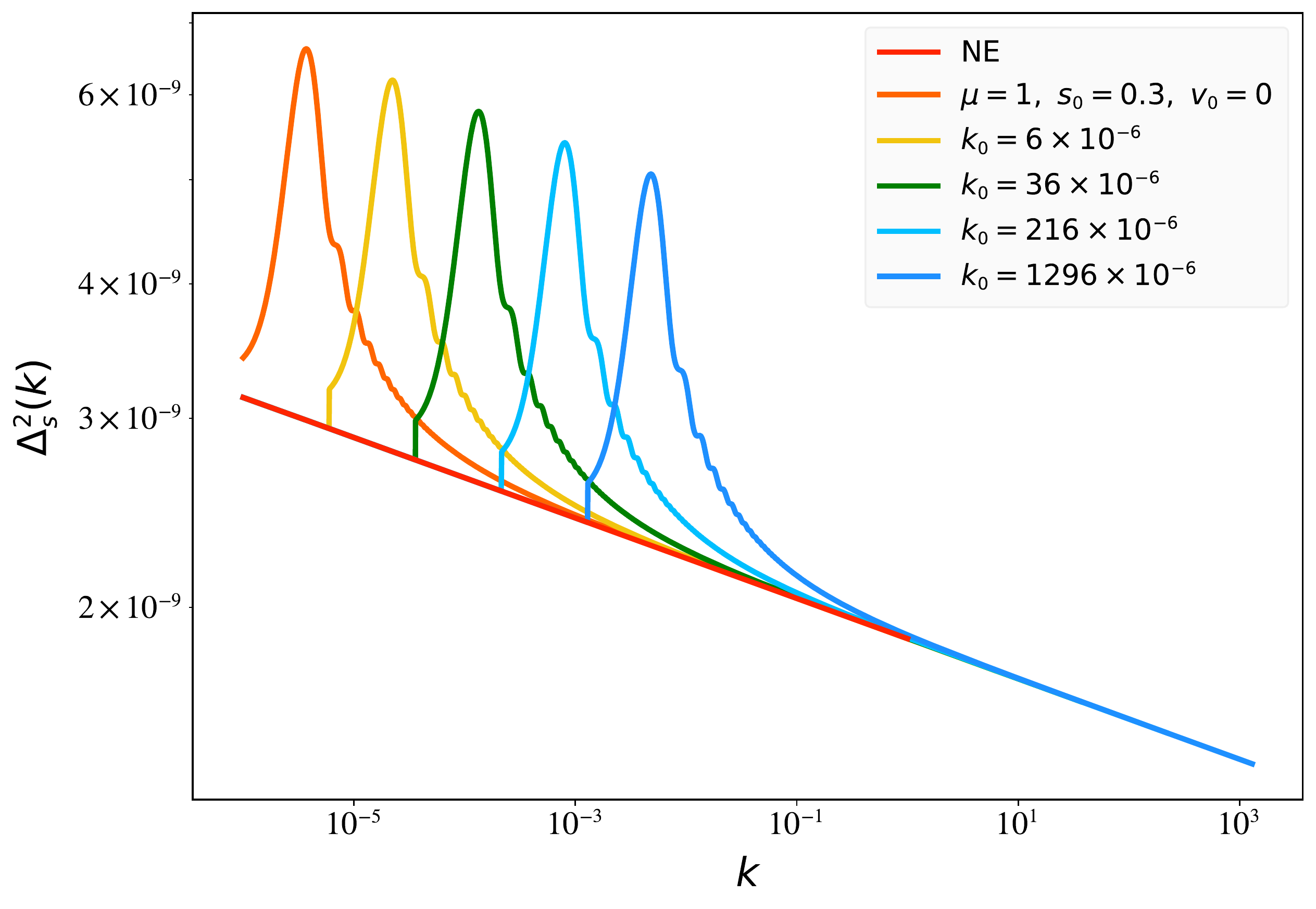}
    \vspace{0.1cm}
  \end{subfigure}
  \begin{subfigure}[b]{0.62\linewidth}
    \includegraphics[width=\linewidth]{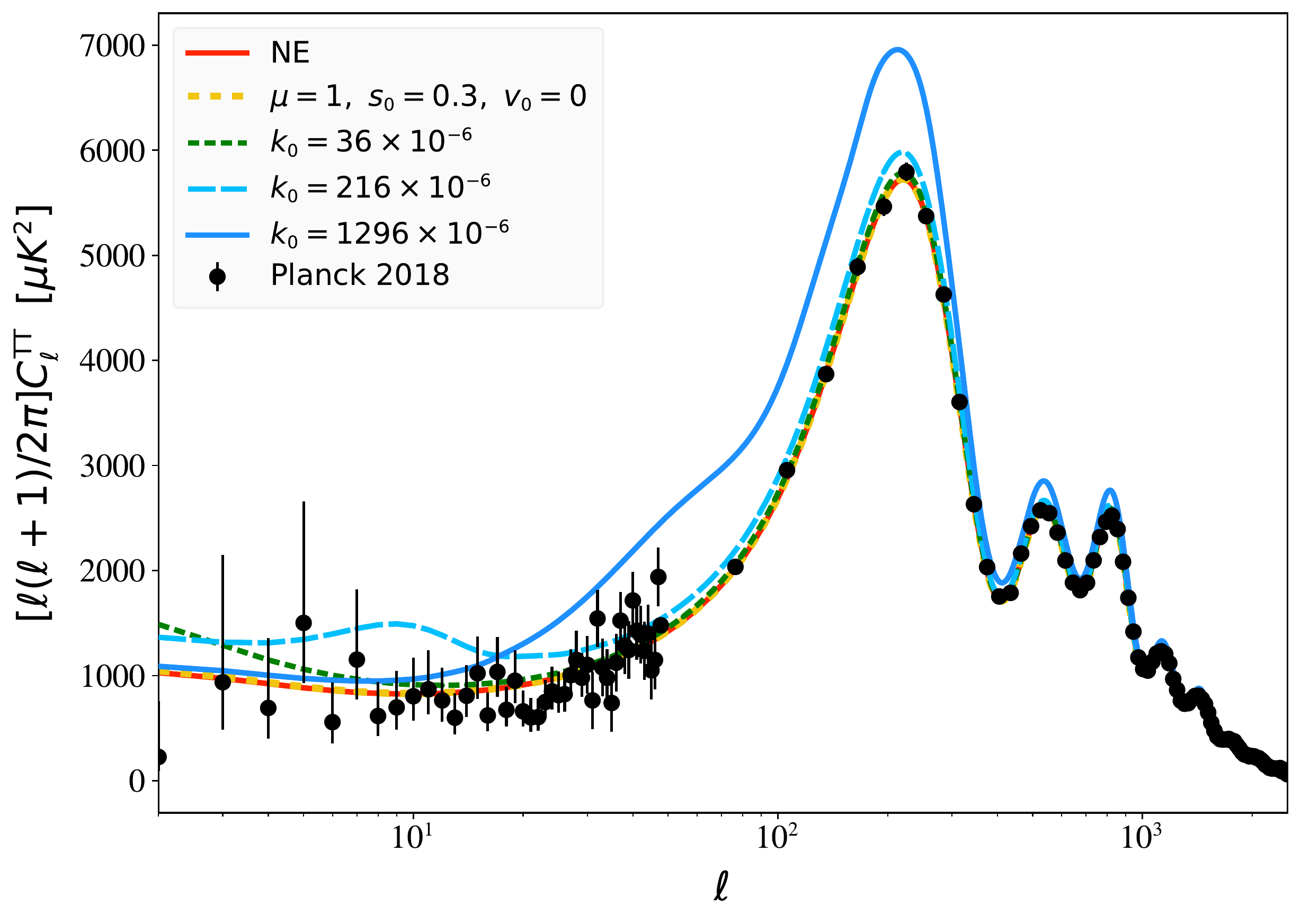}
    \vspace{0.1cm}
  \end{subfigure}
  \begin{subfigure}[b]{0.62\linewidth}
    \includegraphics[width=\linewidth]{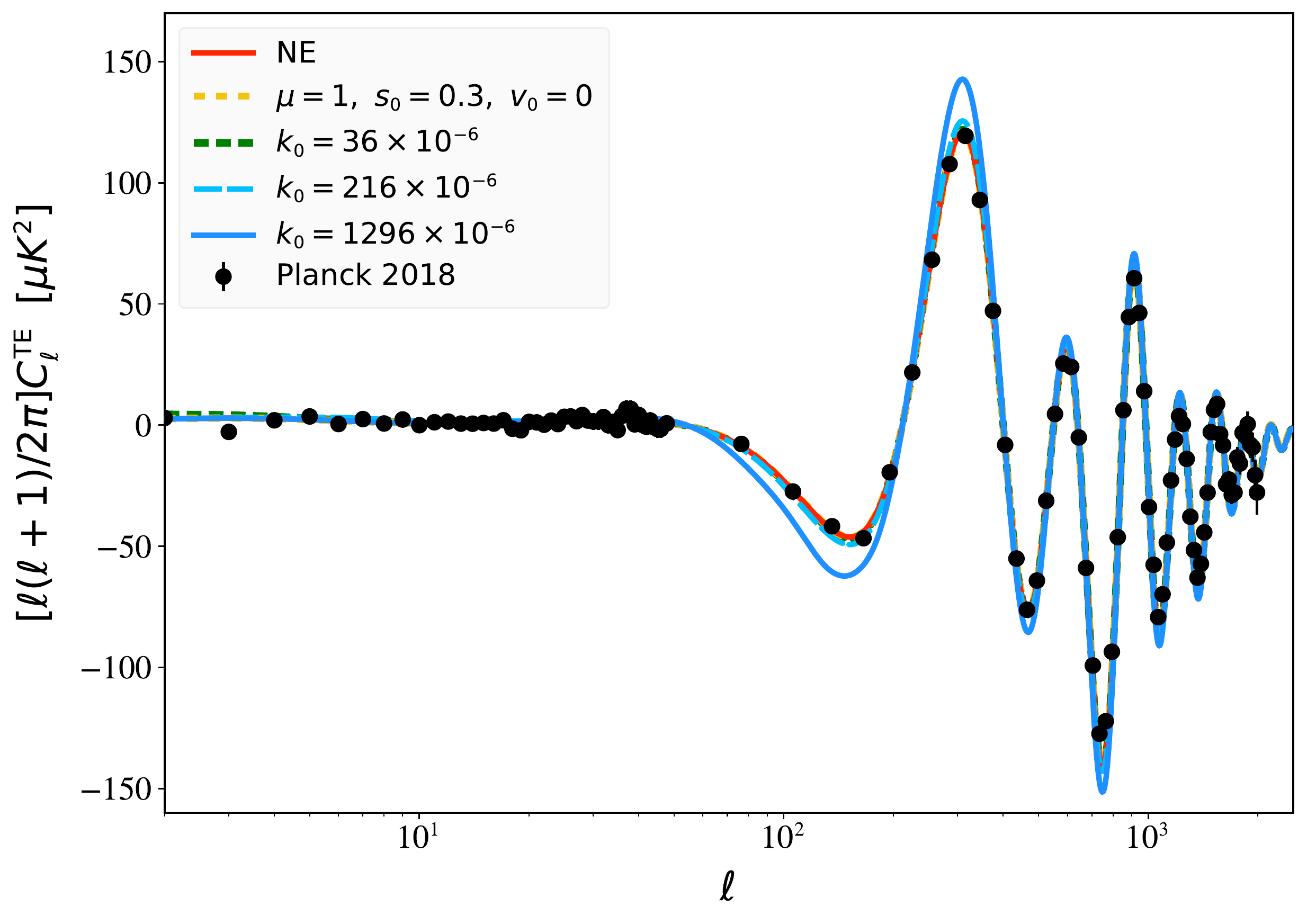}
  \end{subfigure}
\caption{The primordial power spectrum (top), the unlensed TT power spectrum (middle), and the unlensed TE power spectrum (bottom) for an entangled state involving a free massive scalar field with $\mu =1$, $s_{0} = 0.3$, and $v_0 =0$, for various values of $k_{0}$. As in Figure \ref{{fig:ClandPkmu0.1nos0k0shift}}, the non-entangled power spectra are plotted in all subfigures. Furthermore, the Planck CMB data is displayed in the middle and bottom subfigures.}
\label{{fig:ClandPkmu1nov0k0shift}}
\end{figure}

\begin{figure}[hbtp]
\centering
  \begin{subfigure}[b]{0.62\linewidth}
    \includegraphics[width=\linewidth]{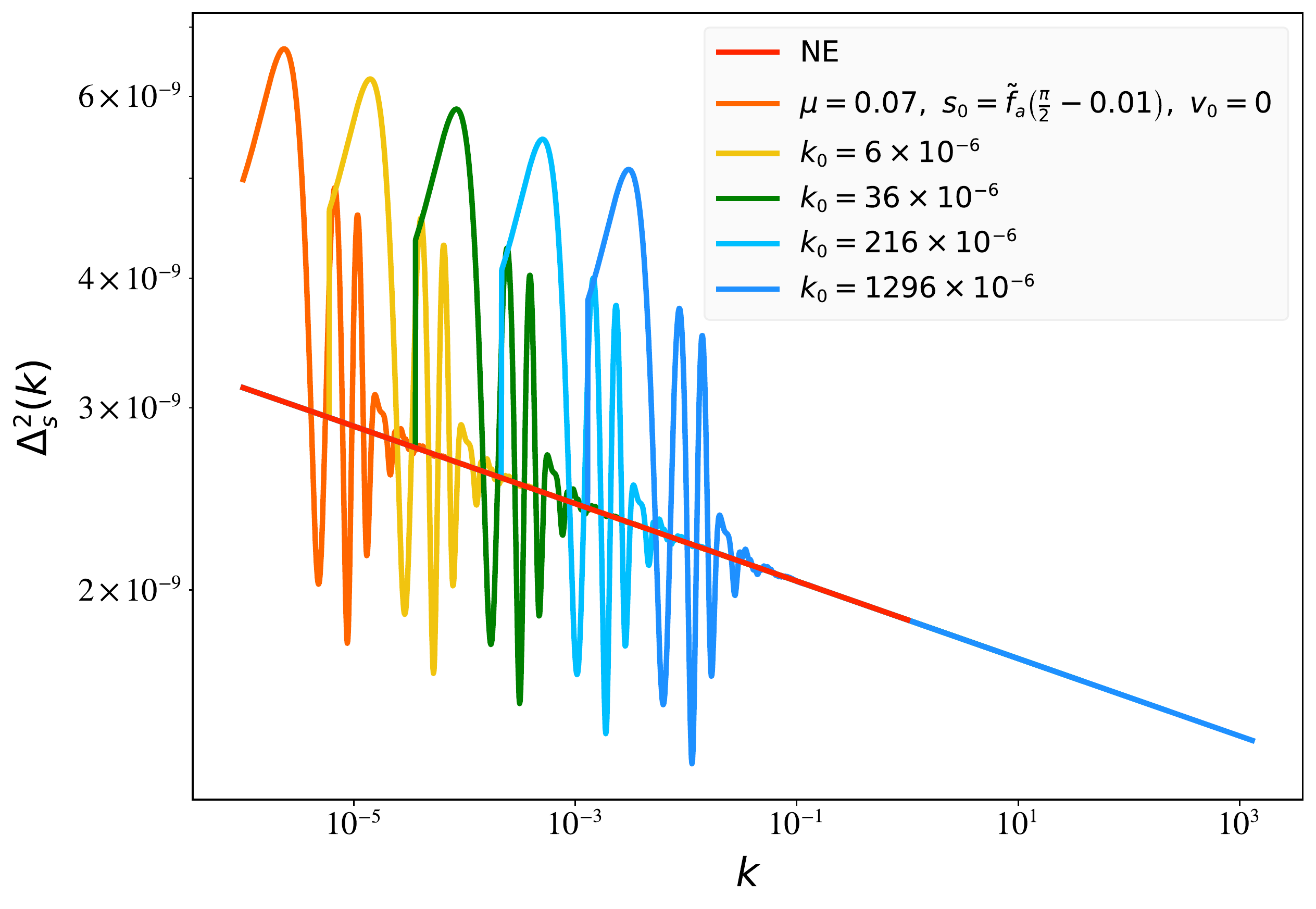}
    \vspace{0.1cm}
  \end{subfigure}
  \begin{subfigure}[b]{0.62\linewidth}
    \includegraphics[width=\linewidth]{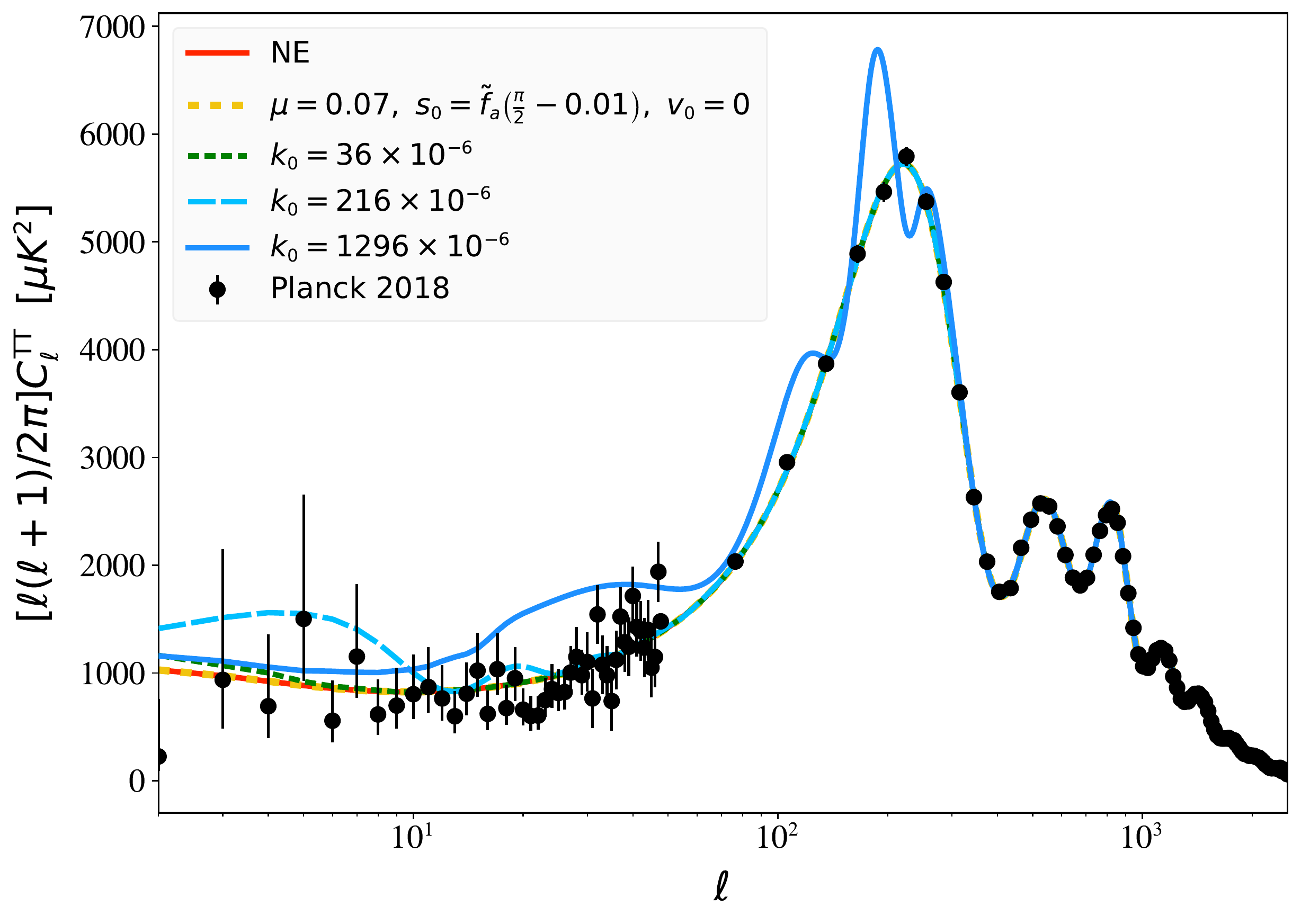}
    \vspace{0.1cm}
  \end{subfigure} 
  \begin{subfigure}[b]{0.62\linewidth}
    \includegraphics[width=\linewidth]{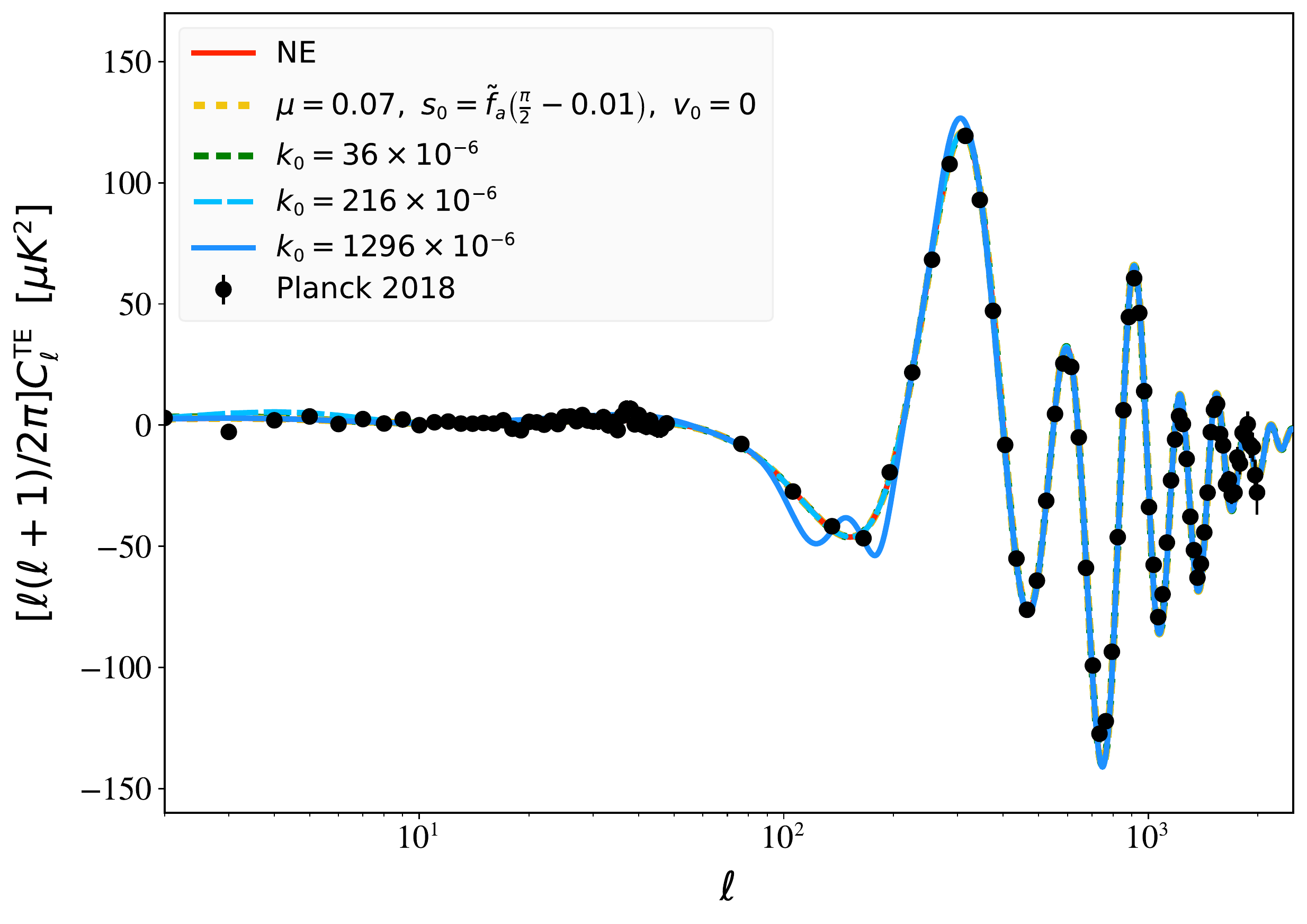}
  \end{subfigure}
\caption{The primordial power spectrum (top), the unlensed TT power spectrum (middle), and the unlensed TE power spectrum (bottom) for an entangled state involving an axion with $\mu=0.07$, $s_0= \tilde{f}_{a} (\frac{\pi}{2} - 0.01)$, and $v_0 =0$, with $f_{decay} = 0.01$, for various values of $k_{0}$. In all subfigures, the non-entangled power spectra are presented. Additionally, the Planck CMB data is shown in the middle and bottom subfigures.}
\label{{fig:ClandPkaxionk0shift}}
\end{figure}

Figure \ref{{fig:ClandPkmu0.1nos0k0shift}} explores shifting the onset of entanglement for the free massive scalar potential with parameters $\mu=0.1$, $s_0= 0$, and $v_0 =0.05$. There are some visual differences in the TT spectra in the low-$l$ regime, particularly for the cases where entanglement starts latest or, equivalently, where $k_{0}$ is the largest. In those situations, the curves dip below the non-entangled TT spectra. Overall, however, the entangled results are similar to what was obtained with this set of parameters in the previous sections.

Next, Figure \ref{{fig:ClandPkmu1nov0k0shift}} does the same comparison for the parameters $\mu=1$, $s_0= 0.3 $, and $v_0 =0$ in the entangled case. These entangled primordial power spectra contain a dominant isolated feature followed by a behavior that approaches the standard non-entangled case. Consequently, the results of shifting the onset of entanglement are more dramatic. In Figure \ref{{fig:ClandPkmu1nov0k0shift}}, one can see the latest onset of entanglement considered---a value of $k_0 = 1296 \times 10^{-6} \; {\rm Mpc}^{-1}$---gives TT and TE spectra outside the bounds of existing data. However, the rest of the shifts considered are much closer to the non-entangled case and most fit the Planck error bars (by eye) just as well as the standard case. The most notable features that are different from the non-entangled case appear in the low-$l$ region of the TT spectra and appear vaguely oscillatory for some values of $k_0$.

Finally, Figure \ref{{fig:ClandPkaxionk0shift}} displays an identical comparison for the axion-like potential, with parameters $s_0= \tilde{f}_{a} (\frac{\pi}{2} - 0.01) $, $\mu=0.07$, $v_0 =0$, and $\tilde{f}_{a} = 0.01$ for the entangled case. Like Figure \ref{{fig:ClandPkmu1nov0k0shift}}, the entangled primordial power spectra contain a dominant isolated feature, followed by behavior that matches the non-entangled case. Furthermore, as in Figure \ref{{fig:ClandPkmu1nov0k0shift}}, the latest onset of entanglement considered produces results that appear beyond the bounds of the Planck error bars. However, the rest of the shifts considered in Figure \ref{{fig:ClandPkaxionk0shift}} show TT spectra that match the standard non-entangled case for $l > 100$, yet have a distinct imprint of damped oscillations (compared to the non-entangled case) for the low-$l$ regime. The amplitude of these low-$l$ oscillations are related to the amplitude of oscillations in the primordial spectrum, so some amount of tuning by adjusting initial parameters is definitely possible. The TE spectra are also a good match for all but the latest onset of entanglement considered.

Overall, Figures \ref{{fig:ClandPkmu0.1nos0k0shift}} - \ref{{fig:ClandPkaxionk0shift}} showcase the effects of changing the onset of entanglement, and they demonstrate how changing this parameter enables one to put features in the CMB power spectra (particularly in the low-$l$ regime) where, in some cases, there previously were none.

\section{Conclusions}
\label{sec:conclusions}

There are some interesting lessons to take away from our analysis. The first is that entangled Gaussian states might be easier to generate than previously thought. The only ingredient necessary is the existence of a scalar field that is displaced from its minimum and/or has an initial velocity. As we mentioned above, there are enough such fields in most extensions of the standard model with these properties. Furthermore, as opposed to the analysis in~\cite{Holman:2019spa} where the cubic $\zeta-\Sigma$ action was considered, even the quadratic action considered here can generate a non-trivial entangled state.

So, can the Planck data distinguish the BD state from one of its entangled analogs? It is clear that, even by eye, some parameter values are excluded due to new features generated in the TT and/or TE spectrum. On the other hand, some seemingly reasonable parameter values seem to fit the data well, again at least by eye. Whether these parameter values can survive the scrutiny of a full parameter estimation probe is future work on this project. We can also make use of bi-spectrum information~\cite{Akrami:2019izv} as in~\cite{Bolis:2019fmq} to further constrain the parameters of the entangled state. 

We note that the changes in the $C_l$s (due to entanglement in the state) seem to be most significant when the time at which entanglement turns on is well within the last 55 e-folds of inflation.  At one level, this result is not surprising; we note that the trend is to settle back down to the non-entangled case after an initial (sometimes large) deviation from it. If this initial deviation happens early enough (i.e., when the largest distance scales appearing in the CMB leave the inflationary horizon), then for most of the subsequent evolution the power spectrum is essentially the standard Bunch-Davies result. Thus we get the largest observational ``bang for the buck'' when new features appear after the largest observed distance scales leave the inflationary horizon.

Our interpretation of what we have done so far is that an entangled state could well be hiding in the Planck data. It remains to be seen what a full attempt at parameter estimation might yield, but---at least by eye---there appears to be a range of parameters $\mu,\ s_0,\ v_0$ that give CMB anistropies consistent with Planck. If these initial results \emph{are} borne out by further analysis, then finer probes of the CMB will have to be developed to distinguish entangled states from the Bunch-Davies case.

As a parting thought we reiterate the importance of understanding the quantum state of the inflaton. Clearly it is of great importance in terms of understanding cosmological measurements. More significantly though, we hope that a better understanding of which states can possibly be \emph{the} consistent inflationary quantum state will be a signpost guiding us to the next layer of physical laws.


\acknowledgments
RB and NB were supported in part by the U.S. Department of Energy, Office of Science, Office of High Energy Physics QuantISED program under Contract No. KA2401032. SM was supported in part by Prof. Katherine Freese through the College of Natural Sciences Sponsored Research Development funding at The University of Texas at Austin. RB and NB would like to thank Andreas Albrecht for fruitful discussions while RH and BR thank Devon Houtz for title and edit suggestions.



\bibliographystyle{unsrt}
\bibliography{ZetaSpecEnt.bib}







\end{document}